\documentclass[jgrga,draft]{agutex}
\usepackage{dcolumn}% Align table columns on decimal point
\usepackage{bm}% bold math
\usepackage{amsmath}
\usepackage{bm}% bold math
\usepackage{amsfonts}
\usepackage[usenames,dvipsnames]{color}

\usepackage[]{graphicx}
\setkeys{Gin}{draft=false}
\authorrunninghead{CASAGRANDE ET AL.}

\titlerunninghead{MULTISCALE COUPLING IN THE EARTH SYSTEM}

\authoraddr{$^{*}$ Corresponding author: Annalisa Molini, Institute Center for Water and Environment (iWater), Masdar Institute of Science and Technology, PO Box 54224, Abu Dhabi, UAE (amolini@masdar.ac.ae) }

\begin{document}
%\linenumbers\relax % Commence numbering lines
%% ------------------------------------------------------------------------ %%
%
%  TITLE
%
%% ------------------------------------------------------------------------ %%

%\title{Wavelet correlation and multiscale coupling\\ in geophysical systems}
%\title{Wavelet correlations reveal multiscale coupling\\ in geophysical systems}
\title{Wavelet correlations to reveal multiscale coupling\\ in geophysical systems}
%\title{Wavelet correlations to diagnose multiscale coupling\\ in geophysical systems}
%
% e.g., \title{Terrestrial ring current:
% Origin, formation, and decay $\alpha\beta\Gamma\Delta$}
%

%% ------------------------------------------------------------------------ %%
%
%  AUTHORS AND AFFILIATIONS
%
%% ------------------------------------------------------------------------ %%

\authors{Erik Casagrande~\altaffilmark{1}, Brigitte Mueller~\altaffilmark{2}, Diego Miralles~\altaffilmark{3,4},
 Dara Entekhabi~\altaffilmark{5} and Annalisa Molini\altaffilmark{1,*}}

\altaffiltext{1}{Institute Center for Water and Environment (iWater), Masdar Institute of Science and Technology, Abu Dhabi, United Arab Emirates.}

\altaffiltext{2}{Environment Canada, M3H 5T4, Toronto, Ontario, Canada.}

\altaffiltext{3}{VU University Amsterdam, Earth and Life Sciences, Amsterdam, The Netherlands.}

\altaffiltext{4}{Ghent University, Laboratory of Hydrology and Water Management, Ghent, Belgium.}

\altaffiltext{5}{Ralph M. Parsons Laboratory for Environmental Science and Engineering, Massachusetts Institute of Technology, Cambridge, Massachusetts, USA.}
\newpage
%% ------------------------------------------------------------------------ %%
%
%  ABSTRACT
%
%% ------------------------------------------------------------------------ %%

% >> Do NOT include any \begin...\end commands within
% >> the body of the abstract.

\begin{abstract}
The interactions between climate and the environment are highly complex. Due to this complexity, process-based models are often preferred to estimate the net magnitude and directionality of interactions in the Earth System. However, these models are based on simplifications of our understanding of nature, thus are unavoidably imperfect. Conversely, observation-based data of climatic and environmental variables are becoming increasingly accessible over large scales due to the progress of space-borne sensing technologies and data-assimilation techniques. Albeit uncertain, these data enable the possibility to start unraveling complex multivariable, multiscale relationships if the appropriate statistical methods are applied.

Here, we investigate the potential of the wavelet cross-correlation method as a tool for identifying multiscale interactions, feedback and regime shifts in geophysical systems. The ability of wavelet cross-correlation to resolve the fast and slow components of coupled systems is tested on synthetic data of known directionality, and then applied to observations to study one of the most critical interactions between land and atmosphere: the coupling between soil moisture and near-ground air temperature. Results show that our method is not only able to capture the dynamics of the soil moisture-temperature coupling over a wide range of temporal scales (from days to several months) and climatic regimes (from wet to dry), but also to consistently identify the magnitude and directionality of the coupling. Consequently, wavelet cross-correlations are presented as a promising tool for the study of multiscale interactions, with the potential of being extended to the analysis of causal relationships in the Earth system.

\end{abstract}

%% ------------------------------------------------------------------------ %%
%
%  BEGIN ARTICLE
%
%% ------------------------------------------------------------------------ %%

% The body of the article must start with a \begin{article} command
%
% \end{article} must follow the references section, before the figures
%  and tables.

\begin{article}

%% ------------------------------------------------------------------------ %%
%
%  TEXT
%
%% ------------------------------------------------------------------------ %%

\section{Introduction}
  \label{Introduction}
Natural systems are often characterized by complex interactions, mainly originating in the overlap of dynamical processes acting at very diverse temporal and spatial scales. Examples of this multiscale dynamics can be found in several branches of geophysics. These include climate and the hydrological cycle, whose different components interact and synchronize over a wide range of scales and patterns \citep{Lovejoy:2013ts,Tsonis:2007ty}, and ecological systems, where resilience and evolution are mainly determined by cooperation and connectivity \citep{Sole:1999wm,Moilanen:2002vj,Cowen:2006ug}. 
In a similar framework it is logical to expect that also the couplings among the different components of the system --  and between the system and the surrounding -- take place at multiple scales and across them. 
Multiscale interactions have recently received extensive attention in the literature, and have been proposed as a mechanism for the triggering of extreme events \citep{Miralles:2014is,Peters:2004wc,Raffa:2008eo}, abrupt regime transitions \citep{Okin:2009he,Peters:2007kk} and patterns formation \citep{Scanlon:2007jg, Guttal:2009uq}. 

Examples of this increasing interest for multiscale and cross-scale interactions can be found in ecology \citep{Allen:2008ts,Moritz:2005,Cash:2006vd,Peters:2007kk,Raffa:2008eo,Scanlon:2007jg,Thrush:2013cv,Werner:2014kp} and climate dynamics \citep{Holbrook:gh,DebraPCPeters:2007cy,Molini:2010ft,Okin:2009he,Rial:2004fz}, but also in fields other than geosciences such as network morphology \citep{Odor:2013fv,PastorSatorras:2001jv} and econometrics \citep{Nikkinen:2011ig}.
Most of these studies are based on minimalist models of interaction across scales~\citep{Allen:2002cq,Peters:2004wc,Peters:2007kk}, or -- when some kind of data-driven approach is attempted -- on classical scaling statistics, more able to resolve the scale-dependent structure of the considered processes, rather than coupling strength and directionality across scales.

Following this approach, the scaling properties of a wide number of dynamical processes including turbulent flows \citep{Frisch:1995wl}, atmospheric tracers such as precipitation \citep{Molini:2010grl,Veneziano:2012eb,Lovejoy:2013ts}, ecosystems patterns and organization \citep{Wu:2006wi,Nagelkerken:2009wc}, social networks \citep{Szell:2010wx}, urban growth and development \citep{Lammer:2006ui,Chen:2008wi,Bettencourt:2007ej,Pumain:2004tt}, and economic systems \citep{Lux:1999fl,Mandelbrot:1998ur,Mandelbrot:1997ur} have been intensively investigated in the last decades through a variety of scaling metrics. 
In contrast, considerably less attention has been devoted to the analysis of couplings and feedbacks taking place at different scales and across them \citep{Dhamala:2008du,Molini:2010ft,Schmitt:2007gl}, to the development of ad-hoc statistics to analyze the temporal evolution of such couplings \citep{Hui:1997,Ngae:1998id,Mizuno:2001,Sello:2000uc,Salvetti:1999vb,Onorato:1997wd,Lungarella:2007,Shirazi:2013}, and to the investigation of the spectral features of systems displaying strong connectivity across inter-annual, seasonal and sub-seasonal scales~\citep[][]{Torrence1998,Torrence1999}.
 
In this study we explore the potential of a simple \emph{local-coupling} metric, the wavelet cross-correlation, for identifying and assessing linear and intermittent interactions across different temporal scales in geophysical systems. 
Time-domain correlations are routinely applied to multivariate geophysical time-series to identify phase consistency among the variables, but they fails when the variability is dominated by periodicity or when the phase consistency is intermittent. Harmonics-based statistics equivalent to the time-domain correlation, like spectral coherence and cross-spectra can partially fill this gap, being able to separate the phase consistency over different persistent frequencies. However, also these spectral statistics are limited in capturing interactions that may be transient in time or may display different directionality across scales.

Conversely, wavelet cross-correlation can be inferred through the wavelet decomposition  of observed signals, and widens the classical concept of multiscale information flow within random multiplicative processes \citep{Arneodo:1998vd}. What is particularly appealing in this measure is its ability to decompose linear correlations in scale and time, preserving simultaneously the total correlation of the system \citep{Daubechies:1992wb}.
Based on the whitening  properties of wavelet filters \citep{Mallat:2008wm,Percival:1999jm}, the wavelet coefficients describing the bi-variate process $\{ {X_t},{Y_t}\}$ at each temporal scale $s$ can be treated as realizations of the jointly Gaussian random process $\{ {{\cal W}_X}(t,s),{{\cal W}_Y}(t,s)\}$ and the sample pairs $({{w}_X}({t_i},s),{{w}_Y}({t_i},s))$  and $({{w}_X}({t_j},s),{{w}_Y}({t_j},s))$ can be considered mutually independent for $i \ne j$. However, if the joint-Gaussianity assumption can in practice be relaxed when we test null cross-correlations, the {\it null auto-correlation} hypothesis suffers from a number of limitations -- mostly depending on the degree of wavelet overlapping at the analyzed scale -- which need to be considered.

The essential theoretical background on wavelet cross-correlation and its significance testing are discussed in section \ref{sec:Methods}, together with some caveats about the inference of directional couplings from spectral statistics.
Section~\ref{autoregressive} is devoted to test the performance of the wavelet cross-correlation in assessing scale-by-scale interaction in synthetic auto-regressive systems of known directional coupling and increasing complexity. 
Following the test on synthetic data, section~\ref{LA} deals with the multiscale nature of land-atmosphere interactions through the example of the air temperature-soil moisture coupling and feedback \citep{Seneviratne:2010jw,Seneviratne:2006gc,Miralles:2012gv,Orlowsky:2010ed}.  
Finally, a further discussion of wavelet cross-correlation strength and criticality is provided in section~\ref{Dis}, together with concluding remarks and future developments.

\section{Methods and Background}
  \label{sec:Methods}
  Wavelet cross-correlations are obtained through the combined use of wavelet filtering \citep{Mallat:2008wm,Daubechies:1992wb} and classic linear coupling measures.\\
  In the time domain, under stationary and ergodic assumptions, the simplest and most adopted measure of linear coupling between the trajectories of a bi-variate real-valued stochastic process $\{ {X_t},{Y_t}\}$ is the cross-correlation function ${\rho _{XY}}(\tau )$ defined as:
\begin{equation}
  \label{eq:crosscorrlag}
  {\rho _{XY}}(\tau ) = \frac{{{\gamma _{XY}}(\tau )}}{{\sqrt {\sigma _X^2\sigma _Y^2} }},
\end{equation}
  %{\rho _{XY}}(\tau ) = \frac{{{C_{XY}}(t,t + \tau )}}{{\sqrt {{C_{XX}}{C_{YY}}} }},
\noindent with $\tau=0,...,k$ representing the time lag (i.e. temporal asymmetry) between the process trajectories, ${\gamma_{XY}}(\tau ) = E[{X_t}{Y_{t + \tau }}]$ their non-centered covariance, and $\sigma^2_X$, $\sigma^2_Y$ the variances of $\{ {X_t}|\cdot \}$ and $\{ \cdot |{Y_t} \}$ respectively. 
Assuming that causes precede effects in time, it is common practice to associate the presence of significant correlations at non-zero lags with causal asymmetric coupling. However, when the analyzed signals display multi-scaling, non-stationarity and periodicity -- or simply some form of oscillatory behavior at different frequencies -- causality can hardly be inferred from classic lagged cross-correlations ~\cite[see][for a detailed discussion of the relationship between causality, co-integration and correlation]{Granger:1969fg}.  
In fact, it is important to note that even if the analyzed signals are expected to display different coupling strengths and synchronization patterns over a wide range of temporal scales -- like often the case in climate and other geophysical systems -- what we finally ``see'' through the estimation of ${\rho _{XY}}(t,t + \tau )$ is only the aggregated effect of these multiscale interactions, that does not necessary reproduce the actual directionality of the coupling at a specific scale $s$.

As an example, interactions between land and atmosphere such as the soil moisture-air temperature coupling discussed in section \ref{LA}, are characterized by strong seasonal synchronization effects that can partially mask the directionality of fine-scale interactions. However, high-frequency components -- through vertical fluxes of energy and water acting at extremely localized spatio-temporal scales -- still play an important role in triggering and sustaining those interactions.
Decomposing the correlation into its scale-by-scale components can hence shade some light on the different dynamical processes characterizing the observed coupling. This can be achieved by simply considering the analogue of lagged correlation in the wavelet domain \citep{Hui:1997,Turbelin:2009us}, i.e. \emph{wavelet cross-correlation}.
  Unlike integral statistics such as wavelet co-spectra and coherence \citep{Grinsted2004,Maraun2004,Torrence1998}, the wavelet cross-correlation is based on the ``a priori'' decomposition of the bi-variate signals through wavelet band-pass filtering and on a direct inference of scale-by-scale linear correlations from the resulting time series of sample coefficients ${{\it w}_{{X_i}}}(t,s)$.

Wavelet cross-correlations were first adopted in a number of experimental studies on multiscale interactions in turbulent flows and mixing \citep{Hui:1997,Ngae:1998id,Sello:2000uc,Salvetti:1999vb,Onorato:1997wd}, mainly based on the continuous wavelet transform (CWT).
%mainly based on continuous wavelet transform. 
Following this bulk of work, Whitcher and coauthors \citep{Whitcher:2000kk,Whitcher:2000fe} proposed an expression for wavelet cross-correlations and their confidence interval based on the maximum overlap discrete wavelet transform (MODWT), a variant without sub-sampling of the orthonormal discrete wavelet transform (DWT) \citep{FoufoulaGeorgiou:1994wx,Percival:1997ve}. The use of MODWT instead of DWT was dictated by the lack of translation invariance in the latter, strongly impacting the final lag-resolution of wavelet cross-correlations. Through the MODWT, it is also possible to reduce the effects of redundancy, partially preserving invariance in the translation \citep{Gencay:2001vqa}. 

However, the trade-off between lag-resolution and redundancy should be carefully evaluated depending on the specific application.
In this study, based on the wide range of temporal scales over which geophysical processes evolve, and on the necessity of retaining as much information as possible about asymmetry in coupling, we opted for using the CWT together with complex analyzing wavelets, known to preserve these properties \citep{Mallat:2008wm}. Also redundancy of CWT cannot be really considered a disadvantage here, until we are not concerned with compression and the effects of auto-correlation at large scales. Rather it can often become an advantage since the redundancy of CWT allows for a better visualization of correlation patterns across scales \citep{Crowley:vt}.
\subsection{Continuous Wavelet Filtering: Generalities}
\label{waveletfiltering}
Let $x(t)$ denote a sample trajectory of the \emph{finite energy} random process $\{{X_t}\}$.
Then via the CWT, we can decompose $x(t)$ into a set of finite basis functions, representing its \emph{variability} at different scales and instants in time ~\citep{Mallat:2008wm,Daubechies:1992wb,Torrence1998}.

The coefficients ${\it w}_X(u,s)$ of the CWT are obtained by decomposing $x(t)$ over dilated and translated  wavelet functions 
$\psi  \in {{\bf{L}}^{\bf{2}}}(\mathbb{R})$ 
of zero average and $\left\| \psi  \right\| = 1$:  
\begin{equation}
\label{eq:cwt}
  {\it w}_X(u,s) = \left\langle {x,{\psi _{u,\,s}}} \right\rangle  = \int_{ - \infty }^{ + \infty } {x(t)\frac{1}{{\sqrt s }}{\psi ^*}\left( {\frac{{t - u}}{s}} \right)dt} 
\end{equation}

\noindent where $s$ is the wavelet scale (inverse of the pseudo-frequency), $u$ is the translation along the time axis, $\psi^*\left(\cdot\right)$ indicates the complex conjugate of the wavelet basis function and $\left\langle \cdot, \cdot \right\rangle $ is the inner product. Since each trajectory of $\{{X_t}\}$ is a sample function from a collection of random variables $\{ {X_{{t_1}}},...,{X_{{t_n}}}\}$ the wavelet filtering process is  asymptotically equivalent to decomposing $\{{X_t}\}$ in a finite number of stochastic processes $\{{\cal W}_X(u,s)\}$ whose realizations are the ${\it w}_X(u,s)$.  
In addition, it is easy to prove that the wavelet transform is equivalent to a convolution with dilated band-pass filters \citep[][p. 79]{Mallat:2008wm}. 

Thus, we can rewrite equation~(\ref{eq:cwt}) as ${{\it w}_X}(u,s) = x \star {\bar \psi _s}(u)$, with ${\bar \psi _s}(u) = {1 \mathord{\left/
 {\vphantom {1 {\sqrt s }}} \right.
 \kern-\nulldelimiterspace} {\sqrt s }}{\psi ^*}\left( {{{ - t} \mathord{\left/
 {\vphantom {{ - t} s}} \right.
 \kern-\nulldelimiterspace} s}} \right)$ and its Fourier transform given by ${{\hat {\bar \psi}}_s}(\omega ) = \sqrt s {\hat \psi ^*}\left( {s\omega } \right)$. It follows that since $\psi$ is zero-average, $\hat \psi (0)$ is also zero and $\hat \psi$ becomes the transfer function of a band-pass filter. 
The CWT extends the benefits of Fourier analysis to observations involving transients and non-stationarities, allowing the estimation of \emph{local} spectral density metrics such as the wavelet scalogram:   
\begin{equation}
  \label{eq:xpower}
  W_X(t,s)=\left|{\it{w}}_X(t,s)\right|^2,
\end{equation}
\noindent and cross-scalogram:
\begin{equation}
  \label{eq:xypower}
  W_{XY}(t,s)={\it w}^*_X(t,s){\it w}_Y(t,s),
\end{equation}
routinely used as a measure of coupling across scales. However, these and other similar metrics such as the wavelet co-spectrum and the wavelet coherence, are not able to provide any information about the temporal asymmetry of couplings at different scales, unlike wavelet cross-correlation metrics. 

\subsection{Multiscale Interactions and Wavelet Cross-correlations}
\label{waveletcorrelation}
We can now define the wavelet covariance of the random process  $\{X_t,Y_t\}$ at lag $\tau$ and scale $s$, as:
\begin{equation}
  \label{eq:WCC}
  \gamma_{XY}(s,\tau) =E[{\cal W}^*_X(t,s){\cal W}_Y(t+\tau,s)]
\end{equation}
%= \mathop {\lim }\limits_{T \to \infty } \frac{1}{T}\int_{ - T/2}^{T/2} {{{\cal W}_x}{{(t,s)}^*}{{\cal W}_y}(t + \tau ,s)dt}.
This can be alternatively expressed as~\citep{Hui:1997}:
\begin{equation}
  \label{eq:WCC2}
{\gamma _{XY}}(s,\tau) = \frac{s}{{2\pi }}\int_{ - \infty }^{ + \infty } {{S_{XY}}(} \omega ){\left| {\hat \psi (s\omega )} \right|^2}{e^{i\tau \omega }}d\omega,
\end{equation}
\noindent where $S_{XY}(\omega )$ is the co-spectrum of the two signals and 
\begin{equation}
  \label{eq:WCC2local}
{\Gamma _{XY}}(\omega ,s) = {S_{XY}}(\omega ){\left| {\hat \psi (s\omega )} \right|^2}
\end{equation}
\noindent is the {\it local wavelet co-spectrum function}. Therefore ${\gamma _{XY}}(s,\tau)$ is the inverse Fourier transform of the local co-spectrum, that integrated across scales gives the classical covariance among the signals ${\gamma _{XY}}(\tau)$.
Although, if the analyzing wavelet $\psi$ is complex also  $\gamma_{XY}(s,\tau)$ is a complex function and can be decomposed into a real part $\Re\left( {{\gamma _{XY}}(s,\tau)} \right)$ and an imaginary part $\Im\left( {{\gamma _{XY}}(s,\tau)} \right)$, bearing information about the strength and the phase of the correlations. It is important to note that in this last case the conservation of ${\gamma _{XY}}(\tau)$ across scales holds only for the real part of the coefficients \citep{Daubechies:1992wb}, so that the corresponding \emph{wavelet cross-correlation} $\rho_{XY}(s,\tau)$ is most commonly estimated as~\citep{Hui:1997}:
\begin{equation}
  \label{eq:WR}
  {\rho _{XY}}(s,\tau) = \frac{{\Re \left( {{\gamma _{XY}}(s,\tau)} \right)}}{{\sqrt {\Re \left( {{{\sigma^2}_X}(s)} \right)\Re \left( {{{\sigma^2}_Y}(s)} \right)} }},
\end{equation}

\noindent where the ${\sigma^2}_i(s)$ represent the variance of the $i$-th variable coefficients at scale $s$. This form of wavelet cross-correlation ranges between $-1$ and $1$ like its time domain counterpart, and represents a simple local decomposition of the cross-correlation function in time. Since the co-variance of a bi-variate complex-valued random process $\{ {{\cal W}_X}(t,s),{{\cal W}_Y}(t,s)\}$ is given by:
\begin{equation}
  \label{eq:complexcov}
  \gamma  = \left( {{\cal W}_X^{\Re}{\kern 1pt} {\cal W}_Y^{\Re} + {\cal W}_X^{\Im}{\kern 1pt} {\cal W}_Y^{\Im}} \right) + i\left( {{\cal W}_X^{\Re}{\kern 1pt} {\cal W}_Y^{\Im} - {\cal W}_X^{\Im}{\kern 1pt} {\cal W}_Y^{\Re}} \right),
\end{equation}
\noindent it is easy to show through symmetry considerations that
\begin{equation}
  \label{eq:WCC3}
  {\Re \left( {{\gamma _{XY}}(s,\tau)} \right)} =2E[\Re \left({\cal W}_X(t,s)\right)\Re \left({\cal W}_Y(t+\tau,s)\right)].
\end{equation}

\begin{figure}[h]
\centering
\noindent\includegraphics[width=40pc]{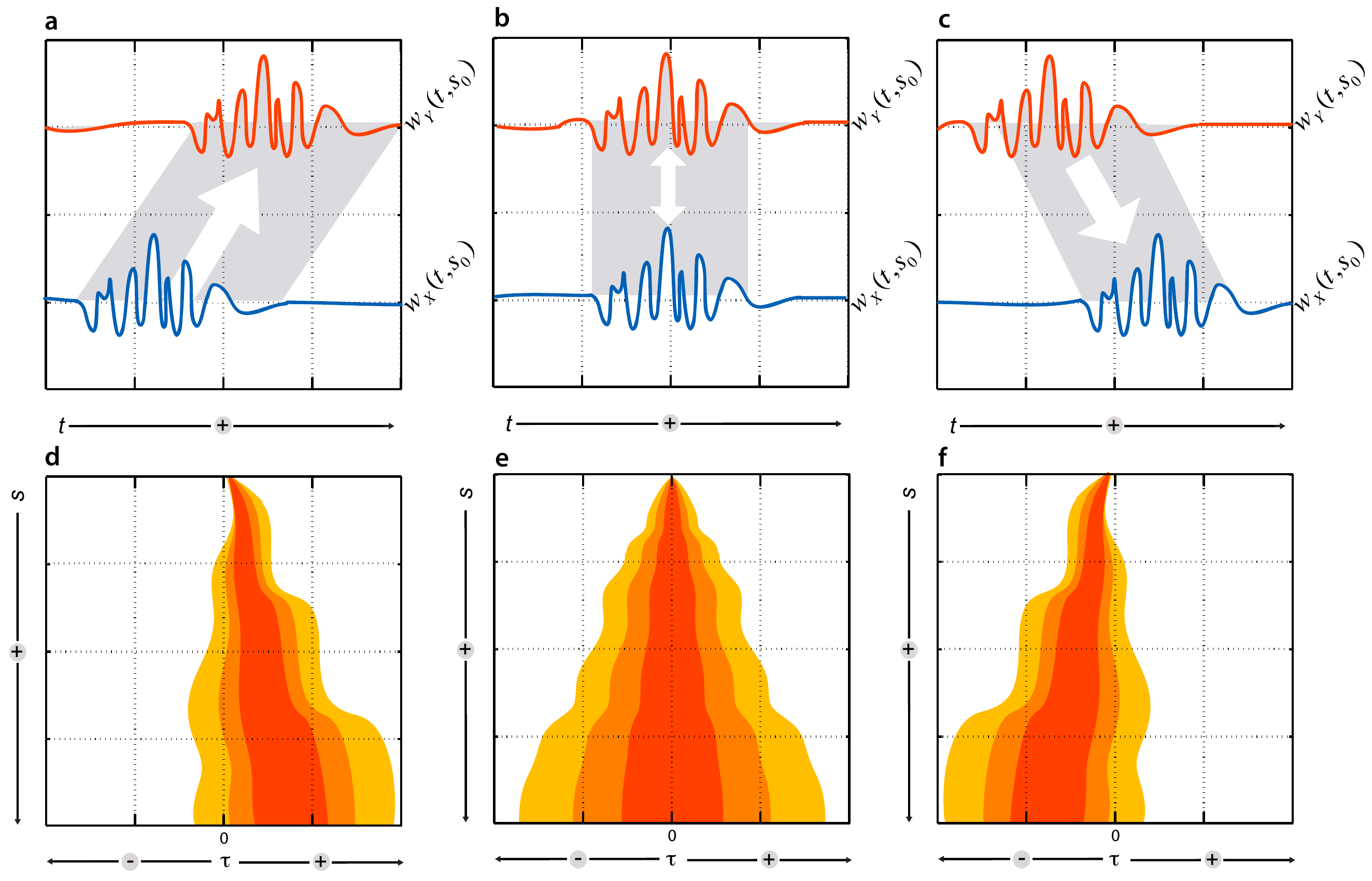}
\caption{Conceptual representation of wavelet coefficients $w_i(t,s_0)$ for $i=X,Y$, as a function of time $t$ and a fixed sample scale $s_0$ (a-c), and wavelet cross-correlation (d-f) for $X$ driving $Y$ (a,d), instantaneous coupling (b,e) and $Y$ forcing $X$ (c,f) for different scales $s$ and time-lags $\tau$ (modified from \citet{Molini:2010ft}). The smallest temporal scale extracted in (d--f) is $s_{1}=\frac{1}{{2{f_\nu }}}$, corresponding to the Nyquist frequency $f_\nu$~\citep[see][for further details]{Torrence1998,Mallat:2008wm,Molini:2010ft}. Forcing direction is taken homogeneous across scales.}
\label{fig:Figure_1}
\end{figure}

Therefore, if an analytic wavelet is used in the decomposition, the asymmetry in the coupling at different scales mainly derives from {\it temporal} shifts among the real parts of the coefficients, rather than from a direct wavelet phase estimation.
Figure~\ref{fig:Figure_1} provides a conceptual representation of wavelet correlation patterns corresponding to $X$ driving $Y$ (d), instantaneous coupling (e) and $Y$ driving $X$ (f) and the associated geometry of sample coefficients at generic scale $s_0$ (a-c). Here the forcing direction is assumed to be homogeneous across scales like in a red-noise dominated process lacking of characteristic forcing scales.

There are a number of alternative formulations for ${\rho_{XY}}(s,\tau)$, such as the one recently proposed by~\citet{Shirazi:2013}, and based on the amplitudes instead of the real part of the coefficients.  
This neglects the information about temporal asymmetries to focus only on the amplitude of correlations across diverse scales. 
\citet{Sello:2000uc} also proposed to estimate scale-by-scale couplings in the form of a local wavelet coherence: 
\begin{equation}
  \label{eq:WLCC}
  {\zeta}_{XY}(s,\tau)=\frac{2\|\gamma_{XY}(\tau,s)\|^2}{\|{{{\sigma^2}_X}(s)}\|^4+\|{{{\sigma^2}_Y}(s)}\|^4}.
\end{equation}

This metric however, varies between $0$ and $1$ due to the absolute operator at the nominator, thus it cannot provide information about the sign of the coupling. 

In the following, ${{\rho }_{XY}}(s,\tau)$ is estimated based on the expression in equation~(\ref{eq:WR}), which is also the one most directly connected with linear couplings in the time domain. The corresponding estimator of ${\rho _{XY}}(s,\tau )$ is then given by:
\begin{equation}
  \label{eq:rhoestimate}
{r_{XY}}(s,\tau )= \frac{{\Re \left( {\sum {w{'^*_X}\left( {i,s} \right)w{'_Y}\left( {i + \tau ,s} \right)} } \right)}}{{\sqrt {\Re \left( {\sum {{{\left[ {w{'_X}\left( {i,s} \right)} \right]}^2}} } \right)\Re \left( {\sum {{{\left[ {w{'_Y}\left( {i,s} \right)} \right]}^2}} } \right)} }},
\end{equation}
\noindent with ${\it w'}\left( {i,s} \right) = {\it w}\left( {i,s} \right) - {\it \bar w}_s$ and ${\it \bar w}_s$ being the long-term average of coefficients at scale $s$.

\subsection{Complex Wavelet Kernels}
\label{kernels}
Different analyzing wavelets can lead to diverse localization effects in frequency and/or in time, that can become crucial for the identification of scale-dependent directional interactions. For this reason, in the following we estimate sample correlations ${r_{XY}}(s,\tau )$ by using two analytic wavelets with different localization in time and scale, i.e. the Morlet and the Paul wavelets. 
The Morlet wavelet is defined as:
\begin{equation}
  \label{eq:morlet}
  {\psi _{{\rm{morl}}}}\left( \eta  \right) = {\pi ^{ - \frac{1}{4}}}{e^{i{\omega _0}\eta  - \frac{1}{2}{\eta ^2}}},
\end{equation}
\noindent while the Paul wavelet is given by:
\begin{equation}
  \label{eq:Paul}
  {\psi _{{\rm{paul}}}}\left( \eta  \right) = \frac{{{2^m}{i^m}m!}}{{\sqrt {\pi (2m)!} }}{(1 - i\eta )^{ - (m + 1)}},
\end{equation}
\noindent where $\eta$ is the non-dimensional time parameter, $\omega_0$ the central frequency of the Morlet wavelet and $m$ the order parameter of the Paul wavelet. 

The Morlet wavelets are more localized (displaying higher resolution) in frequency than in time, while the Paul wavelets display a higher temporal localization~\citep{Torrence1998}. This comparison allows us to address the effects of time/scale resolution on the estimation of the wavelet cross correlations. Following the classic work of~\citet{Torrence1998} we use $\omega_0=6$ and $m=4$ to obtain an effective trade-off in the time and frequency resolution of the decomposition. When working with the Morlet wavelet, the choice of a central frequency $\omega_0=6$ is further justified by the fact that Morlet is not a proper wavelet, as its integral is not zero. However, for $\omega_0$ larger than 5, the integral becomes small enough to ensure the numerical applicability of the Morlet kernel \citep{Kumar:1997um}.
As complex functions, both the Morlet and Paul wavelets are suitable to study oscillatory time series, and have been extensively used in geosciences \citep{FoufoulaGeorgiou:1994wx,Torrence1998}.  
In addition, both the Morlet and Paul wavelets are symmetric, allowing for a non-distorted estimation of temporal shifts. 
This is an important property for robust and reliable analysis of directional couplings.
At the same time, we have to keep in mind that being non-orthogonal, their CWT can be affected by an overlapping of sub-frequency bands and consequent redundancies in the decomposition of the analyzed signal. 

\subsection{Significance Test}
\label{significancetest}
Similar to their time-domain counterparts, multiscale correlations can be tested for significance based on a scale-dependent approach.
In the following, we introduce a simple significance test for wavelet cross-correlations based on the assumption that the coefficients $({{w}_X}({t_i},s),{{w}_Y}({t_i},s))$ are sample pairs drawn from the jointly Gaussian random process $\{ {{\cal W}_X}(t,s),{{\cal W}_Y}(t,s)\}$, and that ${r_{XY}}(s,\tau )$ (${r_{s,t}}$ hereafter for simplicity) is the corresponding sample statistic for the correlation strength and direction at scale $s$. Such an assumption originates from the whitening properties of the wavelet transform~\citep{Mallat:2008wm,Percival:1999jm}, and provides us with the necessary machinery to test ${r_{s,t}}$ against the null correlation hypothesis and construct approximate statistical confidence intervals. We show that in case of {\it{null correlation}} the joint-normality condition can be relaxed owing to the asymptotic properties of the sample correlation distribution in $\rho_{s,\tau}=0$~\citep{Johnson:1995tp}. 

In general, the probability distribution of the sample correlation ${f_R}({r_{s,\,\tau }})$ is cumbersome to derive in a closed-form.
One of the simplest expressions is due to Fisher~\citep{Fisher:1915wo} and is given by:
\begin{align}
\label{eq:fR}
    {f_R}({r_{s,\,\tau }}) = \frac{{{{\left( {1 - \rho _{s,\,\tau }^2} \right)}^{{\textstyle{{n - 1} \over 2}}}}}}{{\pi {\kern 1pt} \Gamma \left( {n - 2} \right)}}{\left( {1 - r_{s,\,\tau }^2} \right)^{{\textstyle{{n - 4} \over 2}}}} \nonumber\\     
     \frac{{{d^{n - 2}}}}{{d{{\left( {{r_{s,\,\tau }}{\rho _{s,\,\tau }}} \right)}^{n - 2}}}}\left\{ {\frac{{{{\cos }^{ - 1}}\left( { - {r_{s,\,\tau }}{\rho _{s,\,\tau }}} \right)}}{{\sqrt {\left( {1 - \rho _{s,\,\tau }^2r_{s,\,\tau }^2} \right)} }}} \right\}   
\end{align}

\noindent where $\Gamma(\cdot)$ is the Gamma function and $n$ the sample size.
Equation~(\ref{eq:fR}), although written in terms of elementary functions, is still too complex to be explicitly used in testing. However, for $\rho_{s,\tau}=0$ it reduces to the null pdf proposed in 1908 by Student~\citep{Kendall:1945vb,Johnson:1995tp}:
\begin{align}
  \label{eq:reduced}
{f_R}({r_{s,\,\tau }})=&\frac{1}{{B\left( {\frac{{n - 2}}{2},\frac{1}{2}} \right)}}{\left( {1 - r_{s,\,\tau }^2} \right)^{\frac{{n - 4}}{2}}} \nonumber\\
=&\frac{{\Gamma \left( {\frac{{n - 1}}{2}} \right)}}{{\Gamma \left( {\frac{1}{2}} \right){\kern 1pt} {\kern 1pt} \,\Gamma \left( {\frac{{n - 2}}{2}} \right)}}{\left( {1 - r_{s,\,\tau }^2} \right)^{\frac{{n - 4}}{2}}},
\end{align}

\noindent where $B(\cdot)$ is the Beta function. Contrary to sample distributions for $\rho  \ne 0$ -- known to be markedly asymmetric -- the probability density function in equation~(\ref{eq:reduced}) is symmetric around $0$ and its derived distribution for   
\begin{equation}
  \label{eq:t1} 
{t_{s,\,\tau }} = \frac{{{r_{s,\,\tau }}}}{{\sqrt {(1 - r_{s,\,\tau }^2)} }}\sqrt {(n - 2)}, 
\end{equation}
\noindent reduces to a t-Student with $(n-2)$ degrees of freedom:
\begin{equation}
  \label{eq:t} 
{f_T}({t_{s,\,\tau }}) = \frac{1}{{\sqrt {(n - 2)} {\kern 1pt} B\left( {\frac{{n - 2}}{2},\frac{1}{2}} \right)}}{\left( {1 + \frac{{t_{s,\tau }^2}}{{n - 2}}} \right)^{ - {\textstyle{{n - 1} \over 2}}}}.
\end{equation}

Therefore, once defined a significance level $\alpha$, equations~(\ref{eq:t1}) and~(\ref{eq:t}) can be used to test wavelet cross-correlations ${r_{s,\tau }}$ against the hypothesis ${H_0}:{\rho _{s,\tau }} = 0$, whose two-tailed rejection region lies outside $\left[ { - {t_{\alpha /2,{\kern 1pt} \,n - 2}},{t_{\alpha /2,{\kern 1pt} \,n - 2}}} \right]$. 
An alternative approach, often adopted in the literature to test the hypothesis $\rho>\rho'$ for $\rho'\neq 0$, relies on a variance-equalizing transformation known as Fisher $z$-transformation:
\begin{equation}
  \label{eq:Zeta} 
{Z_{s,\tau }} = {\tanh ^{ - 1}}{r_{s,\tau }} = \frac{1}{2}\log \left( {\frac{{1 + {r_{s,\tau }}}}{{1 - {r_{s,\tau }}}}} \right),
\end{equation}
\noindent where $Z$ is approximately normally distributed with mean $\mu_z = (1/2) \ln[(1 + \rho_{s,\tau })/(1 - \rho_{s,\tau })]$ and variance $\sigma_z^2 = 1/(n - 3)$. However in this study we limit the approach to test the absence of correlation across scales. 

This choice is motivated by the superior robustness of the $r_{s,\tau}$ estimator in $\rho=0$, and to the possibility of widening the usage of null distribution in equation~(\ref{eq:reduced}) to non jointly-Gaussian samples \citep{Johnson:1995tp}.
In fact, while the expression in equation~(\ref{eq:reduced}) can be seen as an exact representation of ${f_R}$ in $\rho=0$, Fisher's transformation only represents an approximation of ${f_R}$ away from $0$. Consequently, assuming that  linearity and homoscedasticity conditions hold, the ${f_R}$ in $\rho=0$ always  coincides with the null distribution of a jointly-normal process, while the non null distribution of $r$ is robust only under additional kurtosis constrains \citep[see][p.582]{Johnson:1995tp}. 
It is also worth noting that the transformation in equation~(\ref{eq:t}) is generally assumed to be less sensitive to violations of the normality assumption \citep{Edgell:1984vv,Havlicek:1977wx}, i.e. deviations of $\{ {{\cal W}_X}(t,s),{{\cal W}_Y}(t,s)\}$ from a jointly Gaussian distribution should not impact the test for null correlation across scales.

In geosciences it is also a common practice to test spectral statistics -- such as wavelet co-spectra and coherence -- against alternative null hypotheses based on the background noise of the underlying process~\citep{Torrence1999}. The statistical significance at the  desired significance level $\alpha$ is therefore obtained numerically via Monte Carlo simulations. For example, in the case of a red background noise, the null hypothesis is obtained by Monte Carlo replicates~\citep{Torrence1999} through the following steps: (a) a first order autoregressive model of the background noise is fitted to the observed data; (b) a set of surrogates is generated from the fitted model and (c) a suitable confidence interval is computed producing the desired quantile bounds.  However, this numerical approach has been shown to lead to spurious significant correlations in case of reduced samples~\citep{Maraun2004,Maraun2007}. Also, it implies strong assumptions on the background noise of the observed processes.

\section{Wavelet Cross-correlation from Auto-regressive Systems with Pseudo-periodic Features}
\label{autoregressive}
In this section we test the ability of the wavelet cross-correlation $r_{s,\tau}$ to capture multiscale interactions in synthetic processes of known directional coupling.
Large ensembles of numerically generated time series are used to understand whether $r_{s,\tau}$ can be used as an efficient estimator of scale-by-scale coupling strength and directionality for systems of increasing complexity.
\subsection{Coupling in a First-Order Vector Auto-regressive System}\label{AR1}
\begin{figure}[h]
\centering
\noindent\includegraphics[angle=0,width=38pc]{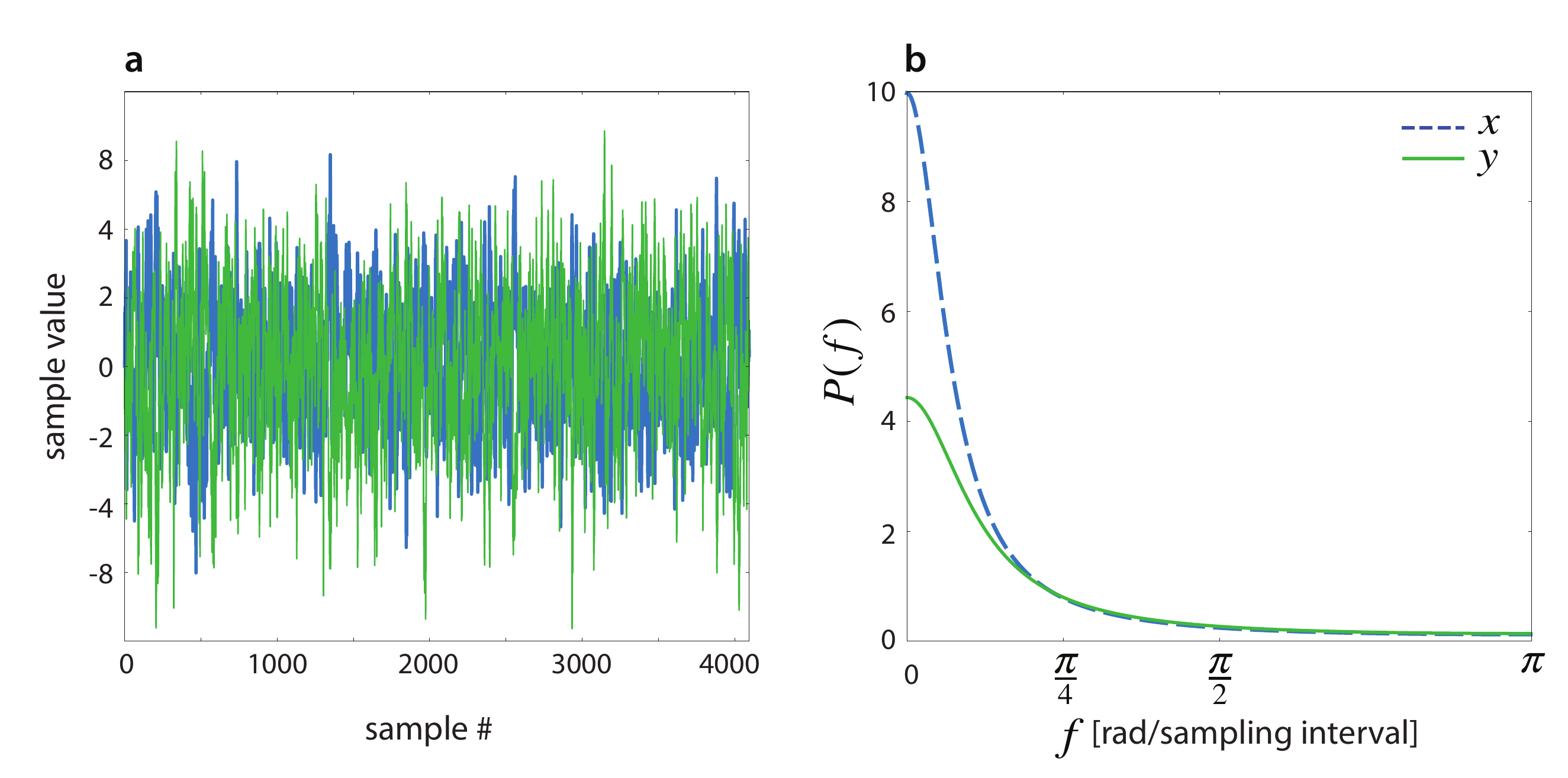}
\caption{Sample realization of the VAR(1) system in equation~(\ref{eq:var1}) (a), and corresponding Lorentzian spectra $P(f)$ (b) for $C_1=-0.4$ and $C_2=0$}
\label{fig:Figure_2}
\end{figure}

The first case study we consider is a simple first order vector auto-regressive model (VAR(1)) in the form:
 \begin{equation}
   \begin{bmatrix}
       x(t) \\
       y(t)
     \end{bmatrix}=
   \begin{bmatrix}0.80& C_1 \\ C_2 &0.70\end{bmatrix}\begin{bmatrix}x(t-1) \\ y(t-1)\end{bmatrix}+\begin{bmatrix}\epsilon(t) \\ \xi(t)\end{bmatrix},
    \label{eq:var1}
 \end{equation}
\noindent where $C_1$ and $C_2$ are coupling parameters defining the strength and directionality of the interaction between $x$ and $y$, and $\epsilon(t)$ and $\xi(t)$ are uncorrelated noise terms with zero mean and unitary variance. 
Figure~\ref{fig:Figure_2}a shows a realization of the VAR(1) model  in equation~(\ref{eq:var1}), where the $x$ and $y$ sub-spaces were linked in a  unidirectional way by imposing a coupling coefficient $C_1=-0.4$ ($y\to x$), and a null feedback from $x$ to $y$  ($C_2=0$). Panel b of the same figure reports the corresponding theoretical power spectra for the $x$ (blue dashed line) and $y$ (green solid line) autoregressive sub-spaces, whose Lorentzian decay is typified by the absence of characteristic scales of autocorrelation.
VAR(1) models, even though simple in construct, still represent a valuable benchmark for $r_{s,\tau}$. 
Red noise, or more in general $1/{f^\alpha }$ noise spectral decay is in fact a characteristic feature of many geophysical systems~\citep{Agnew:1992dj,Keshner:1982fd,Muzzy:2011bw}.
In these models the strength and directionality of the coupling can be easily -- and intuitively -- tuned. Also, the absence of a characteristic scale for the coupling is reflected in ``cascade-like'' $r_{s,\tau}$ patterns similar to the ones sketched in Figure~\ref{fig:Figure_1}.  

\begin{figure}[ht!]
\centering
\noindent\includegraphics[angle=0,width=27pc]{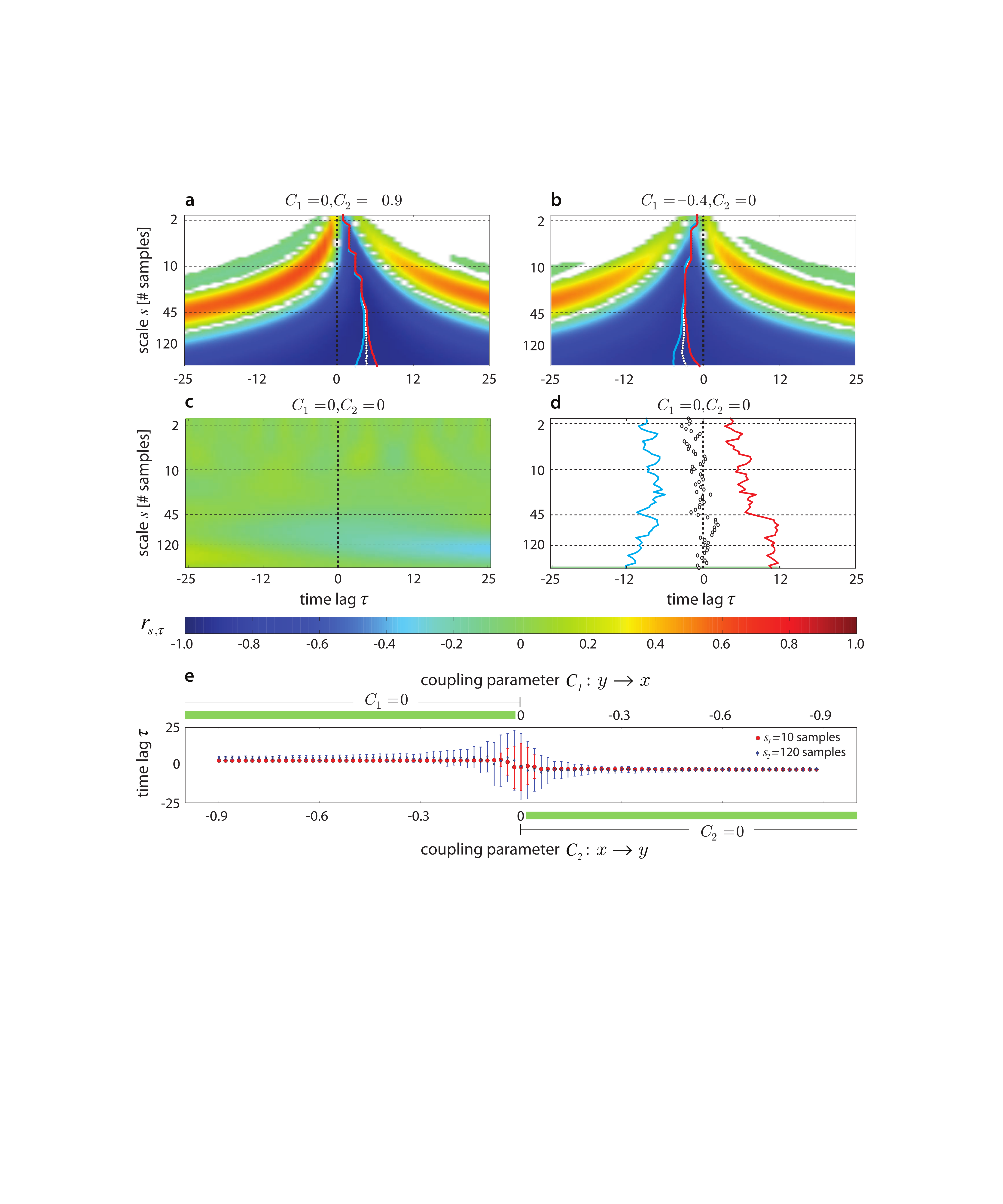}
\caption{Ensemble and ``single-realization'' estimates of $r_{s,\tau}$ for the VAR(1) model in equation~(\ref{eq:var1}) and different coupling parameters $C_1$ and $C_2$. 
Panels a and b respectively depict the ensemble estimates of $r_{s,\tau}$ for $x \to y$ ($C_1=0,C_2=-0.9$) and $y \to x$ ($C_1=-0.4,C_2=0$). Regions with $r_{s,\tau}$ below the $\alpha=99\%$ significance level are masked in white.  The same panels also show $\bar{r}_{min}$ (black empty circles) as defined in section~\ref{AR1} and the corresponding $99\%$ confidence intervals (blue and red solid lines) as a function of scale. 
Wavelet correlation patterns for the uncoupled system ($C_1=C_2=0$) inferred from both a single realization, and an ensemble of simulations are shown in panels c and d, whereas panel e illustrates the sensitivity of $\bar{r}_{min}$ to the coupling strength and directionality for sample scales $s_1=10$ (red dots) and $s_2=120$ (blue diamond) samples.
Bottom and top $x$-axes in (e) respectively represent the coupling strength from from $x$ to $y$ ($C_2$) and $y$ to $x$ ($C_1$), whilst the red and blue bars are the confidence intervals of $\bar{r}_{min}$ for $s=s_1$ and $s=s_2$ as in (a-b) and (d). Note how confidence intervals tend to widen with the weakening of the coupling, and moving from fine to large scales consistently with panels a-b and d.
}
\label{fig:Figure_3}
\end{figure}

Figure~\ref{fig:Figure_3}a-d shows the $r_{s,\tau}$ computed from both single and ensemble-realizations of the VAR(1) model in equation~(\ref{eq:var1}) for different values of the coupling parameters $C_1$ and $C_2$. 
Each ensemble includes $100$ realizations of sample size $n=4096$, independently generated starting from random initial conditions. The wavelet cross-correlation $r_{s,\tau}$ is estimated as an ensemble average over all the $100$ realizations based on a Paul mother wavelet of order $m=4$, and thus represented as a function of temporal asymmetry $\tau$ on the abscissas and scale $s$ on the ordinates. 
Figure~\ref{fig:Figure_3}a depicts the scale-by-scale  correlation patterns resulting from imposing a strong negative coupling from $x$ to $y$ ($C_2=-0.9$) and null feedback from $y$ to $x$ ($C_1=0$),  while Figure~\ref{fig:Figure_3}b shows a case with opposite and weaker coupling, i.e. $C_1=-0.4$ and  $C_2=0$. The considered VAR(1) systems are therefore both negatively coupled but with different strengths, and feedback effects are not considered at this stage. 
Correlations below the $\alpha=99\%$ significance level are masked in white. 

It is interesting to note how in a vector system such as the one in equation~(\ref{eq:var1}), characterized by red-noise-like spectral features and ``short-memory'', the coupling propagates along all the scales in a homogeneous way. To better highlight possible asymmetries in the coupling at different temporal scales $s$ and the variability in asymmetry across the diverse simulations of the ensemble, we also  extracted the minimum correlation $r_{min}=\min \left\{ {{r_{s,\tau}}} \right\}_{\tau =  - k}^k$ for each single realization and derived the ensemble minimum mean $\bar r_{\min}= E[{r_{\min}}]$ and standard deviation ${\sigma _{{r_{\min }}}}$. The focus is here on the minimum of $r_{s,\tau}$ since the two synthetic systems are negatively coupled. However similar considerations on the scale-by-scale maximum of $r_{s,\tau}$ could be done in the case of a positively correlated process.
As discussed next, $\bar r_{\min}$ and ${\sigma _{{r_{\min }}}}$ provide a rough estimate of the asymmetry and variability of peak correlations in the analyzed system. 

In Figure~\ref{fig:Figure_3}a-b,d, $\bar r_{\min}$ (black empty circles) and its confidence interval at the $99\%$ significance level (bounded by the blue and red solid lines) were overlapped to the $r_{s,\tau}$ diagram to show how the peak correlation moves from one side to the other of the $0$ bisecting line when the coupling direction is inverted.
Top panels in Figure~\ref{fig:Figure_3} also show the fast -- quasi exponential -- oscillatory decay  of $r_{s,\tau}$ at different scales $s$, typical of the lagged-cross correlation of an absolutely integrable signal. Depending on the scale, this oscillation decays below the $\alpha=99\%$ significance level more (fine-scales) or less (coarse scales) rapidly, consistently with the fact that the memory of the process is in general weaker at finer scales.
We also test the ability of $r_{s,\tau}$ for identifying null-coupling across-scales by posing $C_1=C_2=0$ in equation~(\ref{eq:var1}). The wavelet cross-correlation patterns obtained from a single-realization of the uncoupled system are shown in Figure~\ref{fig:Figure_3}c,  while panel d depicts the corresponding ensemble statistic. As evident the ensemble $r_{s,\tau}$ in Figure~\ref{fig:Figure_3}d never passes the test for the non-null correlation and can robustly detect the lack of coupling across scales. Also, $\bar r_{\min}$ oscillates around $\tau=0$ with strong variability at all scales.

Figure~\ref{fig:Figure_3}e finally shows the distribution of the minimum value of the  $r_{s,\tau}$ at two selected scales, i.e.~$10$ and $120$ samples, in the case when both coupling and direction are set to vary. Also here, the confidence intervals of $\bar r_{\min}$ are computed from an ensemble of $100$ realizations of the VAR(1) at the level of $99\%$. Panel e shows that for a sufficient strong value of the coupling, $\bar r_{\min}$ can effectively capture the switch in directionality of the coupling at both sample scales. At larger scale ($120$ samples) however, while the value of $\bar r_{\min}$ is higher than for the smaller scale ($10$ samples), the variability is also higher. For any value of the coupling parameter falling in the interval $[C_1>-0.2,C_2>-0.2]$ the estimated range of variability of $\bar r_{\min}$ ends up crossing the zero lag axis and thus does not provide a valid estimate of the directionality of the process. Therefore, the weaker the coupling, the larger the uncertainty on the actual directionality. In addition, uncertainty increases with scale due to the higher  redundancy of wavelet coefficients. 

\subsection{Coupling in a Second-Order Vector Auto-regressive System with Localized Pseudo-periodicity}
\label{AR2}
In this section we test the capability of $r_{s,\tau}$ in detecting couplings that are well-localized in frequency -- i.e. occurring at a specific time-scale. The ability to resolve such localized correlations is important to the study of environmental systems, which often display strong sub-seasonal, seasonal and inter-annual oscillations.
To do so, we consider the following second order vector autoregressive model (VAR(2)) where a {\it pseudo-periodic coupling} from $y$ to $x$ is imposed by choosing roots close to the unit circle:
\begin{equation}
  \label{eq:AR2}
  \begin{bmatrix}x_{t} \\ y_{t}\end{bmatrix}=\begin{bmatrix} 0.55&0.00 \\ 0.00&0.55\end{bmatrix}\begin{bmatrix}x_{t-1} \\ y_{t-1}\end{bmatrix}+\begin{bmatrix} -0.80&-0.30 \\ 0.00&-0.80\end{bmatrix}\begin{bmatrix}x_{t-2} \\ y_{t-2}\end{bmatrix}+\begin{bmatrix}\epsilon_{t} \\ \xi_{t}\end{bmatrix}.
\end{equation}
A similar system is used in~\cite{Dhamala:2008vo} in the context of spectral Granger causality metrics.
Given the localization in frequency of the simulated coupling, the system in equation~(\ref{eq:AR2}) is additionally used to demonstrate the role of wavelet localization in the efficient estimation of $r_{s,\tau}$. 

\begin{figure}[ht!]
\centering
\noindent\includegraphics[angle=0,width=37pc]{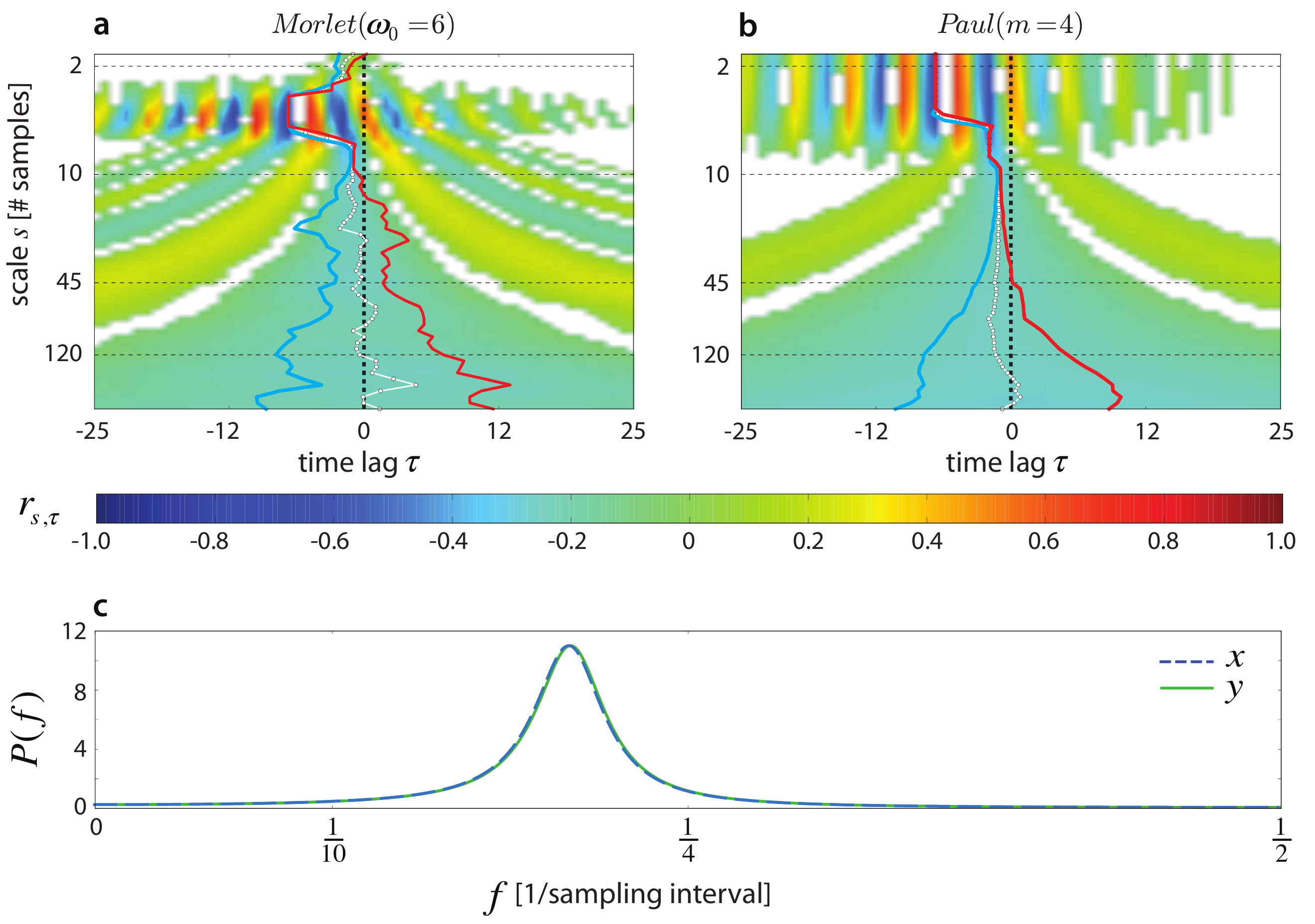}
\caption{Ensemble estimates of $r_{s,\tau}$ and $\bar{r}_{min}$ (same as in (a-b,d) of Figure~\ref{fig:Figure_3}) from the VAR(2) model in equation~(\ref{eq:AR2}), obtained after pass-band filtering with the Morlet (a) and Paul (b) mother wavelets. Panel c shows the corresponding theoretical power spectra $P(f)$ of the auto-regressive sub-spaces of $x$ and $y$. Here, both the Morlet and the Paul wavelets are able to capture the presence of the pronounced periodicity at $1/5$ of of the cycle, in good agreement with the coupling peak shown in panel c for the theoretical spectra. However, Morlet's wavelet, being more localized, can better resolve the coupling peak and its directionality, at the expense of a higher redundancy at large temporal scales.
}
\label{fig:Figure_4}
\end{figure}

Top panels in Figure~\ref{fig:Figure_4} show $r_{s,\tau}$ as estimated from an ensemble of $100$ realizations of equation~(\ref{eq:AR2}), and by respectively using the Morlet (a) and Paul (b) wavelets. Both plots only display values above the $\alpha=99\%$ significance level, and they both show the confidence interval (always $99\%$) for $\bar r_{\min}$ averaged over the $100$ realizations. Panel c reports the theoretical power spectra of each of the $x$ and $y$ autoregressive sub-spaces displaying a clear periodicity at $1/5$ of the cycle.  
Figure~\ref{fig:Figure_4}a-b clearly show that the $r_{s,\tau}$ obtained through the Morlet wavelet better identifies the frequency peak of the coupling in agreement with the characteristics of the kernel, that is by construction, more localized in frequency. In contrast, the frequency peak in the Paul $r_{s,\tau}$ results more spread out, especially in the range of lower scales, but better resolved in time.

Nonetheless, both the Paul and the Morlet $r_{s,\tau}$ are able to correctly detect the directionality ($y \to x$), as evident in the left-asymmetry of $\bar r_{\min}$. It is also worth noting that the variability of $\bar r_{\min}$ is lower where the coupling is stronger (around the frequency peak), gradually increasing with the scale. This is mainly due to the size and relative higher overlapping of the wavelets at larger temporal scales.   
In summary, both the Morlet and the Paul wavelet cross-correlations are able to extract the essential features of the coupling. The choice between the two different wavelets should be made based on the application and the characteristics of the signals under investigation. 

\subsection{Fast/Slow Dynamics Separation and Feedback}
\label{AR2feedback}      
In section~\ref{Introduction} we argue that the evolution of geophysical systems often results from the interaction of diverse dynamical scales. A simple example can be provided by a coupling-feedback system in which the forcing is acting at shorter temporal scales than the response -- i.e. a fast/slow dynamical system.
In such a case, classic correlation analysis in the time domain fails unless the strength of the coupling in one of the two directions is dominant. The wavelet cross-correlation, in contrast, can still resolve the two components of the coupling, their asymmetry and characteristic scales.
Let us consider a modification of the VAR(2) system in equation~(\ref{eq:AR2}) in which $x$ is driving $y$ (negative correlation) at a frequency of $1/15$ of cycle ($f_{slow}$) and $y$ is forcing $x$ (positive correlation) around the $1/6$ of cycle ($f_{fast}$):
\begin{equation}
  \label{eq:AR2bis}
  \begin{bmatrix}x_{t} \\ y_{t}\end{bmatrix}=\begin{bmatrix} 1.73&0.00 \\ 0.00&0.85\end{bmatrix}\begin{bmatrix}x_{t-1} \\ y_{t-1}\end{bmatrix}+\begin{bmatrix} -0.90&-0.10 \\ 0.30&-0.95\end{bmatrix}\begin{bmatrix}x_{t-2} \\ y_{t-2}\end{bmatrix}+\begin{bmatrix}\epsilon_{t} \\ \xi_{t}\end{bmatrix}.
\end{equation}

\begin{figure}[ht!]
\centering
\noindent\includegraphics[angle=0,width=37pc]{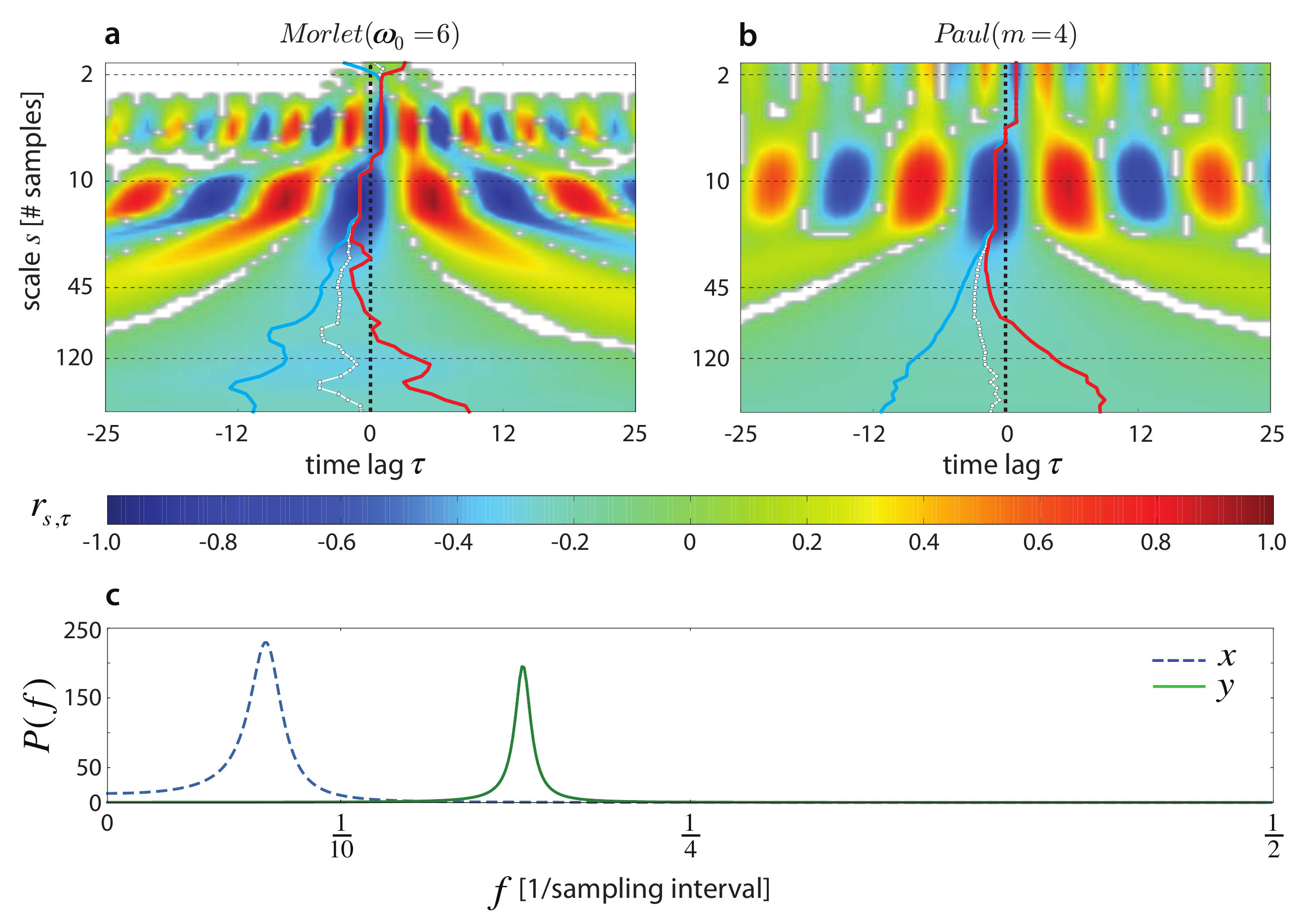}
\caption{Same as Figure~\ref{fig:Figure_4}, but for a system with coupling and feedback occurring at $1/15$ of the cycle and $1/6$ of the cycle respectively, as described in equation~(\ref{eq:AR2bis}).}
\label{fig:Figure_5}
\end{figure}

The corresponding $r_{s,\tau}$ estimated by using both the Morlet and the Paul wavelets is shown in Figure~\ref{fig:Figure_5}, together with the normalized spectra of the two variables. It is evident that also in this case, the $r_{s,\tau}$ is able to resolve the scales at which the coupling actually take place, as well as the opposite asymmetry of the correlation at these scales.
In this case the $\bar r_{\min}$ is only representative of the coupling from $x$ to $y$ due to the negative linear correlation among the two variable at $f_{slow}$. However, mainly as a consequence of the strong periodicity of the coefficients at both the forcing and feedback scales, the localization of $\bar r_{\min}$  is also at some extent representative of the asymmetry in the coupling from $y$ to $x$.
This is overall an interesting result considering that interactions characterized by different degrees of ``memory'' and fluctuation-response relaxation (FRR) effects are ubiquitous in geosciences~\citep{Lacorata2007,Leith1975}.  
\section{Multiscale Interactions in the Soil Moisture-Temperature Coupling}
\label{LA}
Land-atmosphere interactions, their strength and directionality, are one of the main sources of uncertainty in current climate models with strong implications for the accurate assessment of future climate change impacts at regional scales~\citep[see e.g.][]{Seneviratne:2010jw}. Besides the scarcity of direct observations of the states and fluxes across the land-atmosphere continuum, major uncertainties originate from the inherent complexity in the way these variables interact, the multiscale character of these interactions, and the existence of critical tipping points in water and energy availability that may trigger regime transitions. In this last section, we apply the wavelet cross-correlation analysis to a classic form of interaction between land and atmosphere: the coupling between soil moisture ($\theta$) and near-surface air temperature ($T$). 

The objective is to isolate the different components of the coupling across a wide range of temporal scales (from fine weather scales to seasonal to inter-annual) and considering different lag-times between the variables. These two variables ($\theta$, $T$) are mainly related through the process of latent heat flux. They are a priori negatively correlated; however, their coupling can occur in both directions: (a) $T$ may regulate $\theta$ via the drying of the soil due to evaporation, or (b) $\theta$ may regulate $T$ due to evaporative cooling \citep{Seneviratne:2010jw, Miralles:2012gv,Mueller:QVipcOpu}. The latter feedback of $\theta$ on $T$ occurs in regions that are water-limited and it has been referred as a reason why droughts and heatwaves coexist, or follow one another \citep{Quesada:2012jm, Miralles:2014is}.

Nonetheless, the mechanisms through which different dynamical scales contribute to the onset and persistence of the $\theta$-$T$ coupling and feedback remain unclear to a large extent \citep{Orlowsky:2010ed}. In this study we use estimates of daily maximum air temperature obtained from the sub daily screen level (2 m) temperatures of ERA-Interim (http://apps.ecmwf.int/datasets/data/interim full daily/) – the most recent climate reanalysis product of the European Center for Medium Range Weather Forecasts (ECMWF) \citep{Dee:2011ex}. We consider temperature fields for the period 1980-2011 over a global grid with a spatial resolution of 0.75 degrees, corresponding to the native reduced Gaussian grid of ERA \citep{Berrisford:2009uda,Simmons:2014}. Temporal resolution of the assimilated and predicted fields (i.e. the analysis fields) we use here is 6 hours. We extract and analyze maximum daily temperatures as they are most strongly impacted by soil moisture deficits and evaporative cooling effects \citep{Orlowsky:2010ed,Mueller:QVipcOpu}. It is important to note that ERA-Interim temperature is constrained by observations through a complex 4-D assimilation process~\citep{Dee:2011ex}. As a consequence the quality of the analyzed fields is strongly dependent on the density of the station network in the region under consideration \citep[see][]{Simmons:2011vg}. However, the main advantage of reanalysis over direct observations and classical interpolation schemes is their ability to combine observations with a physical model of the atmosphere able to produce physically coherent high-resolution fields and propagate information to areas with poor observational coverage.  For $\theta$ we use an independent data source, the global daily root-zone soil moisture estimates from GLEAM (Global Land-surface Evaporation: the Amsterdam Methodology -- http://foofoo.ugent.be/satex/GLEAM/ as described in \citet{Miralles:2012gv,Miralles:2011hu,Miralles:2011iu,Miralles:2013ix}, also for the period 1980-2011. GLEAM is a set of algorithms designed to retrieve information on evaporation from current satellite observations of hydro-climatic variables~\citep{Miralles:2013ix}. In GLEAM, $\theta$ is derived at daily timescales through the assimilation of satellite soil moisture observations into a multi-layer running water balance that reproduces the infiltration of rainfall through the vertical soil profile \citep{Owe:2008hn}.

\begin{figure}[ht!]
\centering
\noindent\includegraphics[angle=0,width=40pc]{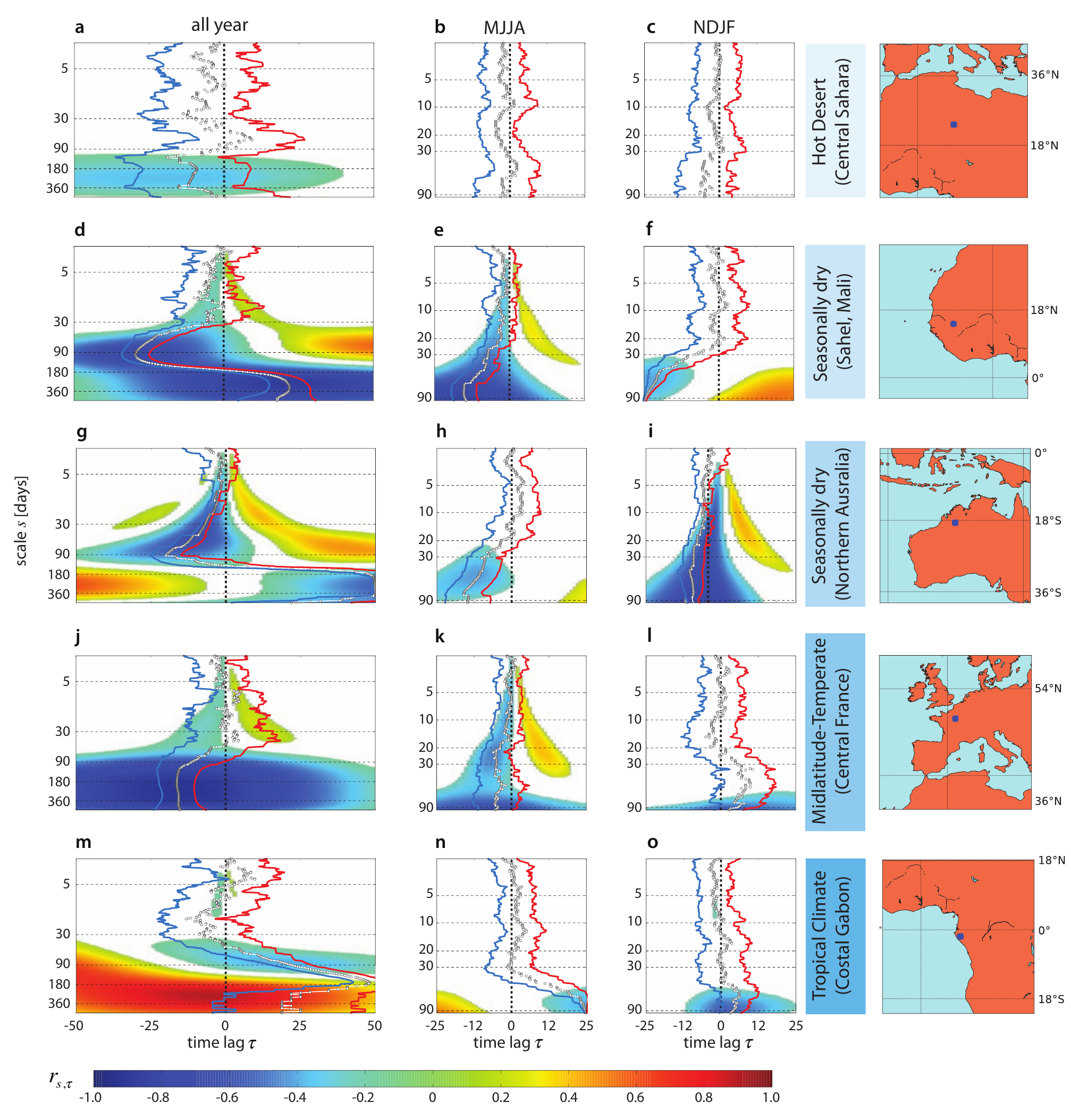}
\caption{Ensemble estimates of $r_{s,\tau}$ and $\bar{r}_{min}$ between soil moisture $\theta$ and air temperature $T$ for five different geographical locations and climatic regimes. Right column shows the exact geographical location of the grid points used in the analysis. Left column reports the ensemble $r_{s,\tau}$ for the full time series at the specified location, while the central columns refers to Boreal summer (MJJA) and winter (NDJF) respectively. From top to bottom, panels are ordered by decreasing aridity. }
\label{fig:Figure_6}
\end{figure}

Figure~\ref{fig:Figure_6}  shows the results of the wavelet cross-correlation analysis for different geographical locations, spanning a wide range of diverse climatic regimes. From top to bottom the panels are ordered by decreasing level of aridity. The panels in the right column show the exact geographical location of the grid points used in the analysis. Wavelet cross-correlations $r_{s,\tau}$ in Figure~\ref{fig:Figure_6}  represent the coupling between $T$ and $\theta$ at a later time, so that $T$ drives $\theta$ if $r_{s,\tau}$ is significant at negative lags and vice versa, $\theta$ drives $T$ for positive lags. The $r_{s,\tau}$ is estimated based on a Paul kernel with m = 4, as an ensemble metric across the available range of years (1980-2011). The resulting correlation patterns can therefore be interpreted as \textit{ensemble averages} of the $\theta$-$T$ coupling across scales. We use here the Paul wavelet based on its superior performance in resolving temporal variability. However, substantially similar results were obtained by using the Morlet kernel. At each location in space, $r_{s,\tau}$  is computed for the entire annual cycle (left panels), only Boreal summer (MJJA, central panels) and only Boreal winter (NDJF, right panels). Before inferring $r_{s,\tau}$, time-series of both $T$ and $\theta$ are normalized to have zero mean and unit variance.  Therefore, harmonic and persistent oscillations are preserved, i.e. we are not specifically analyzing seasonally de-trended anomalies, as one of the main goals is in fact to isolate the relative role of different harmonic components.

As the expected correlation between $T$ and $\theta$ is mostly negative, the minimum correlation ($\bar{r}_{min}$) and its confidence interval can represent a proxy of temporal asymmetry in the multiscale coupling also in this case. The confidence interval of the ensemble seasonal $r_{s,\tau}$ (winter or summer) are computed by considering each summer (or winter) as single time series, resulting in $24$ annual realizations. For the full time series in contrast, the confidence interval is calculated by using a sliding window approach. At each step, $r_{s,\tau}$ is estimated over a time-window of $5$ years, sliding forward with a $1$ year time step. The maximum scale at which $r_{s,\tau}$ can be inferred in the annual and seasonal cases is limited by the size of each sample. Therefore we do not consider any periodicity larger than three months for seasonal samples since this modes are only partially sampled in the data.

Panels a-c show $r_{s,\tau}$ for a location in central Sahara for the full time series (a), MJJA (b), and NDJF (c). Here, the extreme dryness of the soil is expected to inhibit coupling; consequently, a significant correlation can only be found for $T$ driving $\theta$ at the annual scale (circa $360$ days) (see Figure~\ref{fig:Figure_6}a). This coupling, however, is mostly related to the seasonal cycle of temperature since in a hot desert climate, soil moisture retrievals mostly resemble a white noise signal and sensible heat dominates the exchange of energy between land and atmosphere. The extremely erratic behavior of the soil water content is captured by the symmetry of the minimum $\bar{r}_{min}$ (absence of directionality in the coupling) and by the large variability in its confidence interval (Figure~\ref{fig:Figure_6}a). Therefore, hot and dry desert climates can be seen here as a test for null directional coupling across scales, similarly to the synthetic case study reported in Figure~\ref{fig:Figure_3}c-d.

Correlation patterns radically change when we move to a location within the Sahelian sub-region of Mali (Figure~\ref{fig:Figure_6}d-f). In this case most of the precipitation (200-400 mm) falls during the summer months (MJJA), but it is sufficient to trigger the coupling between soil moisture and temperature across a wide range of temporal scales spanning a few days to months (Figure~\ref{fig:Figure_6}d,e). In contrast, the coupling disappears during the winter months (Figure~\ref{fig:Figure_6}f) due to soil desiccation. The seasonally dry character of this climate is captured by the second harmonic at half a year (around 180 days). During the entire year (\ref{fig:Figure_6}d), $T$ leads $\theta$ at time scales of one to 6 months, and $\theta$ leads $T$ at longer time scales.  An analogous pattern can also be identified in Northwestern Australia (Figure~\ref{fig:Figure_6}g-i), another seasonally dry location with a similar separation between humid and water-limited regimes. During Austral summer (NDJF), significant correlations are still confined to the negative lags half-quadrant –- i.e. in average, $T$ drives $\theta$ through evaporation -- although the small scales (few days to 1 month) are clearly the most variable in terms of asymmetry of the coupling. 

This fact could be connected with a higher occurrence of $\theta$ feedbacks on $T$ at these scales throughout the twenty-one years of the ensemble (Figure~\ref{fig:Figure_6}e). Compared to Mali (Figure~\ref{fig:Figure_6}d),  the variability of $\bar{r}_{min}$, and consequently the uncertainty on directionality, is here less pronounced, i.e. the members in the ensemble (g and i) follow very similar correlation patterns at both small and large scales. During the rainy season (Austral winter, Figure~\ref{fig:Figure_6}h) $\theta$ is not limiting evaporation, while during the summer the coupling extends throughout a wide range of scales (Figure~\ref{fig:Figure_6}i), with significant correlations also at positive lags ($\theta$ leading $T$) and at the finest scales \citep{ Dirmeyer:2011ga, Miralles:2012gv}.

The separation between water-limited and \textit{wet} regime is also present in the wavelet cross-correlation patterns of a temperate mid-latitude location in central France (Figure 6j-l). During all year (\ref{fig:Figure_6}j), the coupling is present at larger scales ($>90$ days), and stronger for $T$ leading $\theta$ than $\theta$ leading $T$. During the summer months (\ref{fig:Figure_6}k), negative correlations are found for negative lags, similar to the results for Mali (\ref{fig:Figure_6}i and \ref{fig:Figure_6}e) and indicating an influence of $T$ on $\theta$ at all timescales. On the other hand, during the winter (\ref{fig:Figure_6}l), $\theta$ leads $T$ at large scales. The confidence interval or $\bar{r}_{min}$, however, mostly falls into null correlation regions and thus points to some uncertainties in these results. In general, the seasonal separation is weaker than in Mali (6e and f) and Northwestern Australia (\ref{fig:Figure_6}h and l). Finally, a site located not far from the coast of Gabon is displayed in Figure~\ref{fig:Figure_6}m-o. The site is characterized by a tropical climate and strong moisture advection from the Ocean, hence $\theta$ does not represent a limiting factor in this case, and the only (positive) observed correlation is the one derived from the synchronization between the seasonal cycle of temperature and precipitation in the region \citep[see e.g.,][]{Zhou:2014gl}. 
Overall the wavelet cross-correlation $r_{s,\tau}$  is able to capture the scale-by-scale strength and directionality of the $\theta$-$T$ coupling across different climatic regimes in a consistent way. Moreover, it allows the separation of the local scale contribution from the seasonal signal - often dominant in time domain statistics. Correlation results are stronger in transitional regimes, consistent with previous studies \citep{Koster:2004ge,Seneviratne:2012bn,Miralles:2012gv}, and interestingly, the $\theta$-$T$ coupling propagates across all the analyzed scales during the warm season, when land-atmospheric interactions may be critical for high temperature extremes \citep{Seneviratne:2006gc, Quesada:2012jm, Miralles:2014is}.

\section{Conclusions}
\label{Dis}
We introduced a novel methodology to infer multiscale interactions from observations of dynamical systems that evolve over diverse temporal scales.
The here adopted metric -- the wavelet cross-correlation $r_{s,\tau}$ -- is based on the direct estimation of scale-by-scale correlations in the wavelet domain. 
The ability of $r_{s,\tau}$ to infer interactions across scales -- and their directionality -- was tested on different synthetic coupled systems and on a real-world case study of land-atmosphere interaction the coupling/feedback between soil moisture and near-surface air temperature. When applied to bi-variate auto-regressive vector models of increasing complexity the $r_{s,\tau}$ shows to be able to correctly reproduce the underlying directionality of the coupling at different temporal scales, and to distinguish fast/slow dynamic components within the simulated systems. 
In this context the term directionality is mainly used to indicate some sort of temporally lagged coupling at the considered scale, without any assumption on the causal structure of the observed process.

However, it is clear that the ability of decomposing a coupling in its scale-by-scale components is an attractive feature of wavelet cross-correlation and can be used in disentangle the role of different dynamical processes in coupled geophysical systems. Besides directionality is here mainly a synonymous for temporal asymmetry, and connections between causality and predictability -- like in the case of more proper causality metrics like Granger causality \citep{Granger:1969fg} -- were not yet explored.

The application of wavelet cross-correlations to soil moisture and near surface air 
temperature shows interesting insights into the interaction of these two variables at different climate regimes. An interesting feature of this interaction, when observed through the lens of a multiscale correlation metrics like $r_{s,\tau}$, is that the coupling between soil moisture and temperature, in the passage to a water limited regime, is active through an extremely wide range of scales, and that this feature represents a common signature of the coupling (and possible feedback) across very diverse climatic regimes.
An interesting question arising from this evidence could be whether this activation of the coupling across scales is a feature reproducible in climate models.
A word of caution is finally in order when we consider couplings taking place at large temporal scales ($>$ 1 year), since they could be partially affected by the redundancy -- and consequent auto-correlation effects -- of the continuous wavelet transform.
Overall results of this study demonstrate the potential of wavelet cross-correlations to unravel the relationships between two environmental and climatic variables from a purely statistical perspective. The method here described can, in principle, be applied to observations from any region of the world, and to study soil moisture-temperature coupling or any other multivariate interaction across multiple scales.

%%% End of body of article:

%%%%%%%%%%%%%%%%%%%%%%%%%%%%%%%%
%% Optional Appendix goes here
%
% \appendix resets counters and redefines section heads
% but doesn't print anything.
% After typing \appendix
%
%\section{Here Is Appendix Title}
% will show
% Appendix A: Here Is Appendix Title
%
%%%%%%%%%%%%%%%%%%%%%%%%%%%%%%%%%%%%%%%%%%%%%%%%%%%%%%%%%%%%%%%%
%
% Optional Glossary or Notation section, goes here
%
%%%%%%%%%%%%%%
% Glossary is only allowed in Reviews of Geophysics
% \section*{Glossary}
% \paragraph{Term}
% Term Definition here
%
%%%%%%%%%%%%%%
% Notation -- End each entry with a period.
% \begin{notation}
% Term & definition.\\
% Second term & second definition.\\
% \end{notation}
%%%%%%%%%%%%%%%%%%%%%%%%%%%%%%%%%%%%%%%%%%%%%%%%%%%%%%%%%%%%%%%%
%
%  ACKNOWLEDGMENTS

\begin{acknowledgments}
The data used in this manuscript are publicly available from the ERA Interim (http://apps.ecmwf.int/datasets/data/interim full daily/) and GLEAM (http://foofoo.ugent.be/satex/GLEAM/) online repositories. 

E. Casagrande, D. Entekhabi and A. Molini thankfully acknowledge the funding from Masdar Institute (One-to-One MIT-MI, \#12WAMA1) in the framework of the MIT and Masdar Institute Cooperative Program.
\end{acknowledgments}
\newpage

%% ------------------------------------------------------------------------ %%
%%  REFERENCE LIST AND TEXT CITATIONS

%\bibliography{CSWCbibFinal.bib}

\begin{thebibliography}{95}
\providecommand{\natexlab}[1]{#1}
\expandafter\ifx\csname urlstyle\endcsname\relax
  \providecommand{\doi}[1]{doi:\discretionary{}{}{}#1}\else
  \providecommand{\doi}{doi:\discretionary{}{}{}\begingroup
  \urlstyle{rm}\Url}\fi

\bibitem[{\textit{Agnew}(1992)}]{Agnew:1992dj}
Agnew, D.~C. (1992), {The time-domain behavior of power-law noises},
  \textit{Geophys. Res. Lett.}, \textit{19}(4), 333--336.

\bibitem[{\textit{Allen and Holling}(2002)}]{Allen:2002cq}
Allen, C.~R., and C.~S. Holling (2002), {Cross-scale Structure and Scale Breaks
  in Ecosystems and Other Complex Systems}, \textit{Ecosystems}, \textit{5}(4),
  315--318.

\bibitem[{\textit{Allen and Holling}(2013)}]{Allen:2008ts}
Allen, C.~R., and C.~S. Holling (2013), \textit{{Discontinuities in Ecosystems
  and Other Complex Systems}}, Columbia University Press, New York, USA.

\bibitem[{\textit{Arneodo et~al.}(1998)\textit{Arneodo, Muzy, and
  Sornette}}]{Arneodo:1998vd}
Arneodo, A., J.~F. Muzy, and D.~Sornette (1998),
  {{\textquotedblleft}Direct{\textquotedblright} causal cascade in the stock
  market}, \textit{Eur. Phys. J. B}, \textit{2}, 277--282.

\bibitem[{\textit{Berrisford et~al.}(2011)\textit{Berrisford, Dee, Poli,
  Brugge, Fielding, Fuentes, Kallberg, Kobayashi, Uppala, and
  Simmons}}]{Berrisford:2009uda}
Berrisford, P., D.~Dee, P.~Poli, R.~Brugge, K.~Fielding, M.~Fuentes,
  P.~Kallberg, S.~Kobayashi, S.~Uppala, and A.~Simmons (2011), {The ERA-Interim
  Archive -- Version 2.0}, \textit{Tech. Rep. ERA report series 1, Technical
  Report. ECMWF}, Reading, UK.

\bibitem[{\textit{Bettencourt et~al.}(2007)\textit{Bettencourt, Lobo, Helbing,
  K{\"u}hnert, and West}}]{Bettencourt:2007ej}
Bettencourt, L. M.~A., J.~Lobo, D.~Helbing, C.~K{\"u}hnert, and G.~B. West
  (2007), {Growth, innovation, scaling, and the pace of life in cities.},
  \textit{P. Natl. Acad. Sci. USA}, \textit{104}(17), 7301--7306.

\bibitem[{\textit{Cash et~al.}(2006)\textit{Cash, Adger, Berkes, Garden, Lebel,
  Olsson, Pritchard, and Young}}]{Cash:2006vd}
Cash, D.~W., W.~N. Adger, F.~Berkes, P.~Garden, L.~Lebel, P.~Olsson,
  L.~Pritchard, and O.~Young (2006), {Scale and cross-scale dynamics:
  governance and information in a multilevel world}, \textit{Ecol. Soc.},
  \textit{11}(2), 8.

\bibitem[{\textit{Chen and Zhou}(2008)}]{Chen:2008wi}
Chen, Y., and Y.~Zhou (2008), {Scaling laws and indications of self-organized
  criticality in urban systems}, \textit{Chaos, Solitons {\&} Fractals},
  \textit{35}(1), 85--98.

\bibitem[{\textit{Cowen et~al.}(2006)\textit{Cowen, Paris, and
  Srinivasan}}]{Cowen:2006ug}
Cowen, R.~K., C.~B. Paris, and A.~Srinivasan (2006), {Scaling of connectivity
  in marine populations.}, \textit{Science}, \textit{311}(5760), 522--527.

\bibitem[{\textit{Crowley}(2007)}]{Crowley:vt}
Crowley, P.~M. (2007), {A Guide to Wavelets for Economists}, \textit{J. Econ.
  Surv.}, \textit{21}(2), 207--267.

\bibitem[{\textit{Daubechies}(1992)}]{Daubechies:1992wb}
Daubechies, I. (1992), \textit{{Ten lectures on wavelets}}, SIAM - Society of
  Applied and Industrial mathematics, Philadelphia, USA.

\bibitem[{\textit{{Debra P. C. Peters} et~al.}(2007)\textit{{Debra P. C.
  Peters}, Pielke~Sr, Bestelmeyer, Allen, Munson-McGee, and
  Havstad}}]{DebraPCPeters:2007cy}
{Debra P. C. Peters}, R.~A. Pielke~Sr, B.~T. Bestelmeyer, C.~D. Allen,
  S.~Munson-McGee, and K.~M. Havstad (2007), {Spatial Nonlinearities: Cascading
  Effects in the Earth System}, in \textit{Terrestrial Ecosystems in a Changing
  World}, pp. 165--174, Springer Berlin-Heidelberg, Berlin, Germany.

\bibitem[{\textit{Dee et~al.}(2011)\textit{Dee, Uppala, Simmons, Berrisford,
  Poli, Kobayashi, Andrae, Balmaseda, Balsamo, Bauer, Bechtold, Beljaars,
  van~de Berg, Bidlot, Bormann, Delsol, Dragani, Fuentes, Geer, Haimberger,
  Healy, Hersbach, H{\'o}lm, Isaksen, K{\aa}llberg, K{\"o}hler, Matricardi,
  McNally, Monge-Sanz, Morcrette, Park, Peubey, de~Rosnay, Tavolato,
  Th{\'e}paut, and Vitart}}]{Dee:2011ex}
Dee, D.~P., S.~M. Uppala, A.~J. Simmons, P.~Berrisford, P.~Poli, S.~Kobayashi,
  U.~Andrae, M.~A. Balmaseda, G.~Balsamo, P.~Bauer, P.~Bechtold, A.~C.~M.
  Beljaars, L.~van~de Berg, J.~Bidlot, N.~Bormann, C.~Delsol, R.~Dragani,
  M.~Fuentes, A.~J. Geer, L.~Haimberger, S.~B. Healy, H.~Hersbach, E.~V.
  H{\'o}lm, L.~Isaksen, P.~K{\aa}llberg, M.~K{\"o}hler, M.~Matricardi, A.~P.
  McNally, B.~M. Monge-Sanz, J.~J. Morcrette, B.~K. Park, C.~Peubey,
  P.~de~Rosnay, C.~Tavolato, J.~N. Th{\'e}paut, and F.~Vitart (2011), {The
  ERA-Interim reanalysis: configuration and performance of the data
  assimilation system}, \textit{Q. J. R. Meteorol. Soc.}, \textit{137}(656),
  553--597.

\bibitem[{\textit{Dhamala et~al.}(2008{\natexlab{a}})\textit{Dhamala,
  Rangarajan, and Ding}}]{Dhamala:2008du}
Dhamala, M., G.~Rangarajan, and M.~Ding (2008{\natexlab{a}}), {Analyzing
  information flow in brain networks with nonparametric Granger causality},
  \textit{NeuroImage}, \textit{41}(2), 354--362.

\bibitem[{\textit{Dhamala et~al.}(2008{\natexlab{b}})\textit{Dhamala,
  Rangarajan, and Ding}}]{Dhamala:2008vo}
Dhamala, M., G.~Rangarajan, and M.~Ding (2008{\natexlab{b}}), {Estimating
  Granger causality from fourier and wavelet transforms of time series data.},
  \textit{Phys. Rev. Lett.}, \textit{100}(1), 018701.

\bibitem[{\textit{Dirmeyer}(2011)}]{Dirmeyer:2011ga}
Dirmeyer, P.~A. (2011), {The terrestrial segment of soil moisture-climate
  coupling}, \textit{Geophys. Res. Lett.}, \textit{38}(1), L16702.

\bibitem[{\textit{Edgell and Noon}(1984)}]{Edgell:1984vv}
Edgell, S.~E., and S.~M. Noon (1984), {Effect of violation of normality on the
  \textit{t}-test of the correlation coefficient}, \textit{Psychol. Bull.},
  \textit{95}(3), 576--583.

\bibitem[{\textit{Fisher}(1915)}]{Fisher:1915wo}
Fisher, R.~A. (1915), {Frequency distribution of the values of the correlation
  coefficient in samples from an indefinitely large population},
  \textit{Biometrika}, \textit{10}(4), 507--521.

\bibitem[{\textit{Foufoula-Georgiou and Kumar}(1994)}]{FoufoulaGeorgiou:1994wx}
Foufoula-Georgiou, E., and P.~Kumar (1994), \textit{{Wavelets in Geophysics}},
  Wavelet Analysis and Its Applications, v.4, Academic Press, San Diego, USA.

\bibitem[{\textit{Frisch and Kolmogorov}(1995)}]{Frisch:1995wl}
Frisch, U., and A.~N. Kolmogorov (1995), \textit{{Turbulence}}, The Legacy of
  A. N. Kolmogorov, Cambridge University Press, New York, USA.

\bibitem[{\textit{Gen{\c c}ay et~al.}(2001)\textit{Gen{\c c}ay, Sel{\c c}uk,
  and Whitcher}}]{Gencay:2001vqa}
Gen{\c c}ay, R., F.~Sel{\c c}uk, and B.~J. Whitcher (2001), \textit{{An
  Introduction to Wavelets and Other Filtering Methods in Finance and
  Economics}}, Academic Press, San Diego, USA.

\bibitem[{\textit{Granger}(1969)}]{Granger:1969fg}
Granger, C. W.~J. (1969), {Investigating Causal Relations by Econometric Models
  and Cross-spectral Methods}, \textit{Econometrica}, \textit{37}(3), 424.

\bibitem[{\textit{Grinsted et~al.}(2004)\textit{Grinsted, Moore, and
  Jevrejeva}}]{Grinsted2004}
Grinsted, A., J.~C. Moore, and S.~Jevrejeva (2004), {Application of the cross
  wavelet transform and wavelet coherence to geophysical time series},
  \textit{Nonlinear Proc. Geoph.}, \textit{11}(5-6), 561--566.

\bibitem[{\textit{Guttal and Jayaprakash}(2009)}]{Guttal:2009uq}
Guttal, V., and C.~Jayaprakash (2009), {Spatial variance and spatial skewness:
  leading indicators of regime shifts in spatial ecological systems},
  \textit{Theor. Ecol.}, \textit{2}(1), 3--12.

\bibitem[{\textit{Havlicek and Peterson}(1977)}]{Havlicek:1977wx}
Havlicek, L.~L., and N.~L. Peterson (1977), {Effect of the violation of
  assumptions upon significance levels of the Pearson r}, \textit{Psychol.
  Bull.}, \textit{84}(2), 373--377.

\bibitem[{\textit{Hirschi et~al.}(2010)\textit{Hirschi, Seneviratne,
  Alexandrov, Boberg, Boroneant, Christensen, Formayer, Orlowsky, and
  Stepanek}}]{Hirschi:2010dz}
Hirschi, M., S.~I. Seneviratne, V.~Alexandrov, F.~Boberg, C.~Boroneant, O.~B.
  Christensen, H.~Formayer, B.~Orlowsky, and P.~Stepanek (2010), {Observational
  evidence for soil-moisture impact on hot extremes in southeastern Europe},
  \textit{Nature}, \textit{4}(1), 17--21.

\bibitem[{\textit{Holbrook et~al.}(2014)\textit{Holbrook, Li, Collins,
  Di~Lorenzo, Jin, Knutson, Latif, Li, Power, Huang, and Wu}}]{Holbrook:gh}
Holbrook, N.~J., J.~Li, M.~Collins, E.~Di~Lorenzo, F.-F. Jin, T.~Knutson,
  M.~Latif, C.~Li, S.~B. Power, R.~Huang, and G.~Wu (2014), {Decadal Climate
  Variability and Cross-Scale Interactions: ICCL 2013 Expert Assessment
  Workshop}, \textit{BAMS}, \textit{95}(8), 155--ES158.

\bibitem[{\textit{Johnson et~al.}(1994)\textit{Johnson, Kotz, and
  Balakrishnan}}]{Johnson:1995tp}
Johnson, N.~L., S.~Kotz, and N.~Balakrishnan (1994), \textit{{Continuous
  univariate distributions}}, vol.~2, Wiley-Interscience, New York, USA.

\bibitem[{\textit{Kendall and Stuart}(1945)}]{Kendall:1945vb}
Kendall, M.~G., and A.~Stuart (1945), \textit{{The Advanced theory of
  statistics}}, vol.~I, London, UK.

\bibitem[{\textit{Keshner}(1982)}]{Keshner:1982fd}
Keshner, M.~S. (1982), {1/f noise}, \textit{Proceedings of the IEEE},
  \textit{70}(3), 212--218.

\bibitem[{\textit{Koster et~al.}(2004)\textit{Koster, Dirmeyer, Guo, Bonan,
  Chan, Cox, Gordon, Kanae, Kowalczyk, Lawrence, Liu, Lu, Malyshev, McAvaney,
  Mitchell, Mocko, Oki, Oleson, Pitman, Sud, Taylor, Verseghy, Vasic, Xue,
  Yamada, and Team}}]{Koster:2004ge}
Koster, R.~D., P.~A. Dirmeyer, Z.~C. Guo, G.~Bonan, E.~Chan, P.~Cox, C.~T.
  Gordon, S.~Kanae, E.~Kowalczyk, D.~Lawrence, P.~Liu, C.~H. Lu, S.~Malyshev,
  B.~McAvaney, K.~Mitchell, D.~Mocko, T.~Oki, K.~Oleson, A.~Pitman, Y.~C. Sud,
  C.~M. Taylor, D.~Verseghy, R.~Vasic, Y.~K. Xue, T.~Yamada, and G.~Team
  (2004), {Regions of strong coupling between soil moisture and precipitation},
  \textit{Science}, \textit{305}(5687), 1138--1140.

\bibitem[{\textit{Kumar and Foufoula-Georgiou}(1997)}]{Kumar:1997um}
Kumar, P., and E.~Foufoula-Georgiou (1997), {Wavelet analysis for geophysical
  applications}, \textit{Rev. Geophys.}, \textit{35}(4), 385--412.

\bibitem[{\textit{Lacorata and Vulpiani}(2007)}]{Lacorata2007}
Lacorata, G., and A.~Vulpiani (2007), {Fluctuation-Response Relation and
  modeling in systems with fast and slow dynamics}, \textit{Nonlinear Proc.
  Geoph.}, \textit{14}(5), 681--694.

\bibitem[{\textit{L{\"a}mmer et~al.}(2006)\textit{L{\"a}mmer, Gehlsen, and
  Helbing}}]{Lammer:2006ui}
L{\"a}mmer, S., B.~Gehlsen, and D.~Helbing (2006), {Scaling laws in the spatial
  structure of urban road networks}, \textit{Physica A}, \textit{363}(1),
  89--95.

\bibitem[{\textit{Leith}(1975)}]{Leith1975}
Leith, C.~E. (1975), {Climate Response and Fluctuation Dissipation.},
  \textit{J. Atmos. Sci.}, \textit{32}(1), 2022--2026.

\bibitem[{\textit{Li and Nozaki}(1997)}]{Hui:1997}
Li, H., and T.~Nozaki (1997), {Application of Wavelet Cross-Correlation
  Analysis to a Plane Turbulent Jet.}, \textit{JSME Int. J. B-Fluid T.},
  \textit{40}(1), 58--66.

\bibitem[{\textit{Lovejoy and Schertzer}(2013)}]{Lovejoy:2013ts}
Lovejoy, S., and D.~Schertzer (2013), \textit{{The Weather and Climate:Emergent
  Laws and Multifractal Cascades}}, Cambridge University Press, Cambridge, UK.

\bibitem[{\textit{Lungarella et~al.}(2007)\textit{Lungarella, Pitti, and
  Kuniyoshi}}]{Lungarella:2007}
Lungarella, M., A.~Pitti, and Y.~Kuniyoshi (2007), {Information transfer at
  multiple scales}, \textit{Phys. Rev. E}, \textit{76}(5), 056117.

\bibitem[{\textit{Lux and Marchesi}(1999)}]{Lux:1999fl}
Lux, T., and M.~Marchesi (1999), {Scaling and criticality in a stochastic
  multi-agent model of a financial market}, \textit{Nature},
  \textit{397}(6719), 498--500.

\bibitem[{\textit{Mallat}(2008)}]{Mallat:2008wm}
Mallat, S. (2008), \textit{{A Wavelet Tour of Signal Processing}}, The Sparse
  Way, Academic Press, San Diego, USA.

\bibitem[{\textit{Mandelbrot}(1997)}]{Mandelbrot:1997ur}
Mandelbrot, B.~B. (1997), \textit{{Fractals and Scaling In Finance}},
  Discontinuity, Concentration, Risk, Springer, New York, USA.

\bibitem[{\textit{Mandelbrot and Stewart}(1998)}]{Mandelbrot:1998ur}
Mandelbrot, B.~B., and I.~Stewart (1998), {Fractals and scaling in finance},
  \textit{Nature}, \textit{391}(6669), 758--758.

\bibitem[{\textit{Maraun and Kurths}(2004)}]{Maraun2004}
Maraun, D., and J.~Kurths (2004), {Cross wavelet analysis: significance testing
  and pitfalls}, \textit{Nonlinear Proc. Geoph.}, \textit{11}(4), 505--514.

\bibitem[{\textit{Maraun et~al.}(2007)\textit{Maraun, Kurths, and
  Holschneider}}]{Maraun2007}
Maraun, D., J.~Kurths, and M.~Holschneider (2007), {Nonstationary Gaussian
  processes in wavelet domain: synthesis, estimation, and significance
  testing.}, \textit{Phys. Rev. E}, \textit{75}(1 Pt 2), 016707.

\bibitem[{\textit{Miralles et~al.}(2011{\natexlab{a}})\textit{Miralles, Holmes,
  De~Jeu, Gash, Meesters, and Dolman}}]{Miralles:2011hu}
Miralles, D.~G., T.~R.~H. Holmes, R.~A.~M. De~Jeu, J.~H. Gash, A.~G. C.~A.
  Meesters, and A.~J. Dolman (2011{\natexlab{a}}), {Global land-surface
  evaporation estimated from satellite-based observations}, \textit{Hydrol.
  Earth Syst. Sc.}, \textit{15}, 453--469.

\bibitem[{\textit{Miralles et~al.}(2011{\natexlab{b}})\textit{Miralles, De~Jeu,
  Gash, Holmes, and Dolman}}]{Miralles:2011iu}
Miralles, D.~G., R.~A.~M. De~Jeu, J.~H. Gash, T.~R.~H. Holmes, and A.~J. Dolman
  (2011{\natexlab{b}}), {Magnitude and variability of land evaporation and its
  components at the global scale}, \textit{Hydrol. Earth Syst. Sc.},
  \textit{15}, 967--981.

\bibitem[{\textit{Miralles et~al.}(2012)\textit{Miralles, van~den Berg,
  Teuling, and de~Jeu}}]{Miralles:2012gv}
Miralles, D.~G., M.~J. van~den Berg, A.~J. Teuling, and R.~A.~M. de~Jeu (2012),
  {Soil moisture-temperature coupling: A multiscale observational analysis},
  \textit{Geophys. Res. Lett.}, \textit{39}(2), L21707.

\bibitem[{\textit{Miralles et~al.}(2013)\textit{Miralles, van~den Berg, Gash,
  Parinussa, de~Jeu, Beck, Holmes, Jim{\'e}nez, Verhoest, Dorigo, Teuling, and
  Johannes~Dolman}}]{Miralles:2013ix}
Miralles, D.~G., M.~J. van~den Berg, J.~H. Gash, R.~M. Parinussa, R.~A.~M.
  de~Jeu, H.~E. Beck, T.~R.~H. Holmes, C.~Jim{\'e}nez, N.~E.~C. Verhoest, W.~A.
  Dorigo, A.~J. Teuling, and A.~Johannes~Dolman (2013), {El
  Ni{\~n}o{\textendash}La Ni{\~n}a cycle and recent trends in continental
  evaporation}, \textit{Nature Climate Change}, \textit{4}(2), 122--126.

\bibitem[{\textit{Miralles et~al.}(2014)\textit{Miralles, Teuling, van
  Heerwaarden, and de~Arellano}}]{Miralles:2014is}
Miralles, D.~G., A.~J. Teuling, C.~C. van Heerwaarden, and J.~V.-G. de~Arellano
  (2014), {Mega-heatwave temperatures due to combined soil desiccation and
  atmospheric heat accumulation}, \textit{Nature Geoscience}, \textit{7}(5),
  345--349.

\bibitem[{\textit{Mizuno-Matsumoto et~al.}(2001)\textit{Mizuno-Matsumoto,
  Yoshimine, Nii, Kato, Taniguchi, Lee, Ko, Date, Tamura, Shimojo, Shinosaki,
  Inouye, and Takeda}}]{Mizuno:2001}
Mizuno-Matsumoto, Y., T.~Yoshimine, Y.~Nii, A.~Kato, M.~Taniguchi, J.~K. Lee,
  T.~S. Ko, S.~Date, S.~Tamura, S.~Shimojo, K.~Shinosaki, T.~Inouye, and
  M.~Takeda (2001), {Landau-Kleffner Syndrome: Localization of Epileptogenic
  Lesion Using Wavelet- Cross-Correlation Analysis.}, \textit{Epilepsy Behav.},
  \textit{2}(3), 288--294.

\bibitem[{\textit{Moilanen and Nieminen}(2002)}]{Moilanen:2002vj}
Moilanen, A., and M.~Nieminen (2002), {Simple Connectivity Measures in Spatial
  Ecology}, \textit{Ecology}, \textit{83}(4), 1131.

\bibitem[{\textit{Molini et~al.}(2010{\natexlab{a}})\textit{Molini, Katul, and
  Porporato}}]{Molini:2010ft}
Molini, A., G.~G. Katul, and A.~Porporato (2010{\natexlab{a}}), {Causality
  across rainfall time scales revealed by continuous wavelet transforms},
  \textit{J. Geophys. Res. Atmos.}, \textit{115}(D14), D14123.

\bibitem[{\textit{Molini et~al.}(2010{\natexlab{b}})\textit{Molini, Katul, and
  Porporato}}]{Molini:2010grl}
Molini, A., G.~G. Katul, and A.~Porporato (2010{\natexlab{b}}), {Scale-wise
  evolution of rainfall probability density functions fingerprints the rainfall
  generation mechanism}, \textit{Geophys. Res. Lett.}, \textit{37}(7), L07403.

\bibitem[{\textit{Moritz et~al.}(2005)\textit{Moritz, Morais, Summerell,
  Carlson, and Doyle}}]{Moritz:2005}
Moritz, M.~A., M.~E. Morais, L.~A. Summerell, J.~M. Carlson, and J.~Doyle
  (2005), {Wildfires, complexity, and highly optimized tolerance.}, \textit{P.
  Natl. Acad. Sci. USA}, \textit{102}(50), 17,912--17,917.

\bibitem[{\textit{Mueller and Seneviratne}(2012)}]{Mueller:QVipcOpu}
Mueller, B., and S.~I. Seneviratne (2012), {Hot days induced by precipitation
  deficits at the global scale}, \textit{P Natl. Acad. Sci. USA},
  \textit{109}(31), 12,398--12,403.

\bibitem[{\textit{Muzzy et~al.}(2011)\textit{Muzzy, Bacry, and
  Arneodo}}]{Muzzy:2011bw}
Muzzy, J.~F., E.~Bacry, and A.~Arneodo (2011), {The Multifractal Formalism
  Revised with Wavelets}, \textit{Int. J. Bifurcat. Chaos}, \textit{4}(2),
  245--302.

\bibitem[{\textit{Nagelkerken}(2009)}]{Nagelkerken:2009wc}
Nagelkerken, I. (2009), \textit{{Ecological Connectivity among Tropical Coastal
  Ecosystems}}, Springer Science {\&} Business Media, Dordrecht, The
  Netherlands.

\bibitem[{\textit{Ngae et~al.}(1998)\textit{Ngae, Grignon, and
  Poloniecki}}]{Ngae:1998id}
Ngae, P., M.~Grignon, and J.-G. Poloniecki (1998), {D{\'e}termination d'une
  vitesse de convection {\`a} partir de la d{\'e}composition en ondelettes d'un
  champ thermique}, \textit{Int. J. Therm. Sci.}, \textit{38}(4), 331--339.

\bibitem[{\textit{Nikkinen et~al.}(2011)\textit{Nikkinen, Pynn{\"o}nen, Ranta,
  and V{\"a}h{\"a}maa}}]{Nikkinen:2011ig}
Nikkinen, J., S.~Pynn{\"o}nen, M.~Ranta, and S.~V{\"a}h{\"a}maa (2011),
  {Cross-dynamics of exchange rate expectations: a wavelet analysis},
  \textit{Int. J. Financ. Econ.}, \textit{16}(3), 205--217.

\bibitem[{\textit{{\'O}dor}(2013)}]{Odor:2013fv}
{\'O}dor, G. (2013), {Spectral analysis and slow spreading dynamics on complex
  networks}, \textit{Phys. Rev. E}, \textit{88}(3), 032109.

\bibitem[{\textit{Okin et~al.}(2009)\textit{Okin, Parsons, Wainwright, Herrick,
  Bestelmeyer, Peters, and Fredrickson}}]{Okin:2009he}
Okin, G.~S., A.~J. Parsons, J.~Wainwright, J.~E. Herrick, B.~T. Bestelmeyer,
  D.~C. Peters, and E.~L. Fredrickson (2009), {Do Changes in Connectivity
  Explain Desertification?}, \textit{BioScience}, \textit{59}(3), 237--244.

\bibitem[{\textit{Onorato et~al.}(1997)\textit{Onorato, Salvetti, and
  Buresti}}]{Onorato:1997wd}
Onorato, M., M.~V. Salvetti, and G.~Buresti (1997), {Application of a wavelet
  cross-correlation analysis to DNS velocity signals}, \textit{Eur. J. Mech.,
  B/Fluids}, \textit{16}, 575--597.

\bibitem[{\textit{Orlowsky and Seneviratne}(2010)}]{Orlowsky:2010ed}
Orlowsky, B., and S.~I. Seneviratne (2010), {Statistical Analyses of
  Land-Atmosphere Feedbacks and Their Possible Pitfalls}, \textit{J. Climate},
  \textit{23}(14), 3918--3932.

\bibitem[{\textit{Owe et~al.}(2008)\textit{Owe, de~Jeu, and
  Holmes}}]{Owe:2008hn}
Owe, M., R.~de~Jeu, and T.~Holmes (2008), {Multisensor historical climatology
  of satellite-derived global land surface moisture}, \textit{J Geophys. Res.
  Earth}, \textit{113}(F1), F01002.

\bibitem[{\textit{Pastor-Satorras and
  Vespignani}(2001)}]{PastorSatorras:2001jv}
Pastor-Satorras, R., and A.~Vespignani (2001), {Epidemic Spreading in
  Scale-Free Networks}, \textit{Phys. Rev. Lett.}, \textit{86}(14), 3200--3203.

\bibitem[{\textit{Percival}(1999)}]{Percival:1999jm}
Percival, D.~B. (1999), {Wavelet-based surrogates for testing time series}, in
  \textit{Proceedings of the First Joint BMES/EMBS Conference (BMEEMB-99)},
  vol.~2, p. 910, IEEE.

\bibitem[{\textit{Percival and Mofjeld}(1997)}]{Percival:1997ve}
Percival, D.~B., and H.~O. Mofjeld (1997), {Analysis of Subtidal Coastal Sea
  Level Fluctuations Using Wavelets}, \textit{J. Am. Stat. Assoc.},
  \textit{92}(439), 868--880.

\bibitem[{\textit{Peters et~al.}(2007)\textit{Peters, Bestelmeyer, and
  Turner}}]{Peters:2007kk}
Peters, D., B.~Bestelmeyer, and M.~Turner (2007), {Cross{\textendash}Scale
  Interactions and Changing Pattern{\textendash}Process Relationships:
  Consequences for System Dynamics}, \textit{Ecosystems}, \textit{10}(5),
  790--796.

\bibitem[{\textit{{Peters, Debra P. C.} et~al.}(2004)\textit{{Peters, Debra P.
  C.}, Pielke, Bestelmeyer, Allen, Munson-McGee, and Havstad}}]{Peters:2004wc}
{Peters, Debra P. C.}, R.~A. Pielke, B.~T. Bestelmeyer, C.~D. Allen,
  S.~Munson-McGee, and K.~M. Havstad (2004), {Cross-scale interactions,
  nonlinearities, and forecasting catastrophic events.}, pp. 15,130--15,135.

\bibitem[{\textit{Pumain}(2004)}]{Pumain:2004tt}
Pumain, D. (2004), {Scaling laws and urban systems}, \textit{SFI Working
  Paper}, pp. 2004--02--002.

\bibitem[{\textit{Quesada et~al.}(2012)\textit{Quesada, Vautard, Yiou, Hirschi,
  and Seneviratne}}]{Quesada:2012jm}
Quesada, B., R.~Vautard, P.~Yiou, M.~Hirschi, and S.~I. Seneviratne (2012),
  {Asymmetric European summer heat predictability from wet and dry southern
  winters and springs}, \textit{Nature Climate Change}, \textit{2}(10),
  736--741.

\bibitem[{\textit{Raffa et~al.}(2008)\textit{Raffa, Aukema, Bentz, Carroll,
  Hicke, Turner, and Romme}}]{Raffa:2008eo}
Raffa, K.~F., B.~H. Aukema, B.~J. Bentz, A.~L. Carroll, J.~A. Hicke, M.~G.
  Turner, and W.~H. Romme (2008), {Cross-scale Drivers of Natural Disturbances
  Prone to Anthropogenic Amplification: The Dynamics of Bark Beetle Eruptions},
  \textit{BioScience}, \textit{58}(6), 501--517.

\bibitem[{\textit{Rial et~al.}(2004)\textit{Rial, Pielke~Sr, Beniston,
  Claussen, Canadell, Cox, Held, de~Noblet-Ducoudr{\'e}, Prinn, Reynolds, and
  Salas}}]{Rial:2004fz}
Rial, J.~A., R.~A. Pielke~Sr, M.~Beniston, M.~Claussen, J.~Canadell, P.~Cox,
  H.~Held, N.~de~Noblet-Ducoudr{\'e}, R.~Prinn, J.~F. Reynolds, and J.~D. Salas
  (2004), {Nonlinearities, Feedbacks and Critical Thresholds within the Earth's
  Climate System}, \textit{Climatic Change}, \textit{65}(1-2), 11--38.

\bibitem[{\textit{Salvetti et~al.}(1999)\textit{Salvetti, Beux, and
  Lombardi}}]{Salvetti:1999vb}
Salvetti, M.~V., F.~Beux, and G.~Lombardi (1999), {Application of a Wavelet
  Cross-Correlation Technique to the Analysis of Mixing}, \textit{AIAA
  Journal}, \textit{37}(8), 1007--1010.

\bibitem[{\textit{Scanlon et~al.}(2007)\textit{Scanlon, Caylor, Levin, and
  Rodriguez-Iturbe}}]{Scanlon:2007jg}
Scanlon, T.~M., K.~K. Caylor, S.~A. Levin, and I.~Rodriguez-Iturbe (2007),
  {Positive feedbacks promote power-law clustering of Kalahari vegetation},
  \textit{Nature}, \textit{449}(7159), 209--212.

\bibitem[{\textit{Schmitt and Chainais}(2007)}]{Schmitt:2007gl}
Schmitt, F.~G., and P.~Chainais (2007), {On causal stochastic equations for
  log-stable multiplicative cascades}, \textit{Eur. Phys. J. B},
  \textit{58}(2), 149--158.

\bibitem[{\textit{Sello and Bellazzini}(2000)}]{Sello:2000uc}
Sello, S., and J.~Bellazzini (2000), {Wavelet Cross-Correlation Analysis of
  Turbulent Mixing from Large-Eddy-Simulations},
  \textit{arXiv:physics/0003029v1}, p. 3029.

\bibitem[{\textit{Seneviratne and Koster}(2012)}]{Seneviratne:2012bn}
Seneviratne, S.~I., and R.~D. Koster (2012), {A Revised Framework for Analyzing
  Soil Moisture Memory in Climate Data: Derivation and Interpretation},
  \textit{Journal of Hydrometeorology}, \textit{13}(1), 404--412.

\bibitem[{\textit{Seneviratne et~al.}(2006)\textit{Seneviratne, L{\"u}thi,
  Litschi, and Sch{\"a}r}}]{Seneviratne:2006gc}
Seneviratne, S.~I., D.~L{\"u}thi, M.~Litschi, and C.~Sch{\"a}r (2006),
  {Land{\textendash}atmosphere coupling and climate change in Europe},
  \textit{Nature}, \textit{443}(7108), 205--209.

\bibitem[{\textit{Seneviratne et~al.}(2010)\textit{Seneviratne, Corti, Davin,
  Hirschi, Jaeger, Lehner, Orlowsky, and Teuling}}]{Seneviratne:2010jw}
Seneviratne, S.~I., T.~Corti, E.~L. Davin, M.~Hirschi, E.~B. Jaeger, I.~Lehner,
  B.~Orlowsky, and A.~J. Teuling (2010), {Investigating soil
  moisture{\textendash}climate interactions in a changing climate: A review},
  \textit{Earth-Sci. Rev.}, \textit{99}(3-4), 125--161.

\bibitem[{\textit{Shirazi et~al.}(2013)\textit{Shirazi, Aghamohammadi, Anvari,
  Bahraminasab, Rahimi~Tabar, Peinke, Sahimi, and Marsili}}]{Shirazi:2013}
Shirazi, A.~H., C.~Aghamohammadi, M.~Anvari, A.~Bahraminasab, M.~R.
  Rahimi~Tabar, J.~Peinke, M.~Sahimi, and M.~Marsili (2013), {Scale dependence
  of the directional relationships between coupled time series}, \textit{J.
  Stat. Mech.}, \textit{2}, P02042.

\bibitem[{\textit{Simmons}(2011)}]{Simmons:2011vg}
Simmons, A. (2011), {From observations to service delivery: Challenges and
  opportunities}, \textit{WMO Bulletin}, \textit{60}(2).
  
\bibitem[{\textit{Simmons et~al.}(2014)\textit{Simmons, Poli, Dee, Berrisford,
  Hersbach, Kobayashi, and Peubey}}]{Simmons:2014}
Simmons, A.~J., P.~Poli, D.~P. Dee, P.~Berrisford, H.~Hersbach, S.~Kobayashi,
  and C.~Peubey (2014), {Estimating low-frequency variability and trends in
  atmospheric temperature using ERA-Interim}, \textit{Q. J. R. Meteorol. Soc.},
  \textit{140}(679), 329--353.  

\bibitem[{\textit{Sol{\'e} et~al.}(1998)\textit{Sol{\'e}, Manrubia, Benton,
  Kauffman, and Bak}}]{Sole:1999wm}
Sol{\'e}, R.~V., S.~C. Manrubia, M.~Benton, S.~Kauffman, and P.~Bak (1998),
  {Criticality and scaling in evolutionary ecology}, \textit{Trends Ecol.
  Evol.}, \textit{14}(4), 156--160.

\bibitem[{\textit{Szell et~al.}(2010)\textit{Szell, Lambiotte, and
  Thurner}}]{Szell:2010wx}
Szell, M., R.~Lambiotte, and S.~Thurner (2010), {Multirelational organization
  of large-scale social networks in an online world.}, \textit{P. Natl. Acad.
  Sci. USA}, \textit{107}(31), 13,636--13,641.

\bibitem[{\textit{Thrush et~al.}(2013)\textit{Thrush, Hewitt, Lohrer, and
  Chiaroni}}]{Thrush:2013cv}
Thrush, S.~F., J.~E. Hewitt, A.~M. Lohrer, and L.~D. Chiaroni (2013), {When
  small changes matter: the role of cross-scale interactions between habitat
  and ecological connectivity in recovery}, \textit{Ecol. Appl.},
  \textit{23}(1), 226--238.

\bibitem[{\textit{Torrence and Compo}(1998)}]{Torrence1998}
Torrence, C., and G.~P. Compo (1998), {A practical guide to wavelet analysis},
  \textit{B. Am. Meteorol. Soc.}, \textit{79}(1), 61--78.

\bibitem[{\textit{Torrence and Webster}(1999)}]{Torrence1999}
Torrence, C., and P.~J. Webster (1999), {Interdecadal changes in the
  ENSO-monsoon system}, \textit{J. Climate}, \textit{12}(8), 2679--2690.

\bibitem[{\textit{Tsonis and Elsner}(2007)}]{Tsonis:2007ty}
Tsonis, A.~A., and J.~B. Elsner (2007), \textit{{Nonlinear Dynamics in
  Geosciences}}, Springer Science {\&} Business Media, Dordrecht, The
  Netherlands.

\bibitem[{\textit{Turbelin et~al.}(2008)\textit{Turbelin, Ngae, and
  Grignon}}]{Turbelin:2009us}
Turbelin, G., P.~Ngae, and M.~Grignon (2008), {Wavelet cross-correlation
  analysis of wind speed series generated by ANN based models}, \textit{Renew.
  Energ.}, \textit{34}(4), 1024--1032.

\bibitem[{\textit{Veneziano and Lepore}(2012)}]{Veneziano:2012eb}
Veneziano, D., and C.~Lepore (2012), {The scaling of temporal rainfall},
  \textit{Water Resour. Res.}, \textit{48}(8), W08516.

\bibitem[{\textit{Werner et~al.}(2014)\textit{Werner, Davis, Skelly, Relyea,
  Benard, and McCauley}}]{Werner:2014kp}
Werner, E.~E., C.~J. Davis, D.~K. Skelly, R.~A. Relyea, M.~F. Benard, and S.~J.
  McCauley (2014), {Cross-Scale Interactions and the Distribution-Abundance
  Relationship}, \textit{PLOS ONE}, \textit{9}(5), e97387.

\bibitem[{\textit{Whitcher and Jensen}(2000)}]{Whitcher:2000kk}
Whitcher, B., and M.~J. Jensen (2000), {Wavelet estimation of a local long
  memory parameter}, \textit{Explor. Geophys.}, \textit{31}(2), 94.

\bibitem[{\textit{Whitcher et~al.}(2000)\textit{Whitcher, Guttorp, and
  Percival}}]{Whitcher:2000fe}
Whitcher, B., P.~Guttorp, and D.~B. Percival (2000), {Wavelet analysis of
  covariance with application to atmospheric time series}, \textit{J. Geophys.
  Res. Atmos.}, \textit{105}(D11), 14941.

\bibitem[{\textit{Wu}(2006)}]{Wu:2006wi}
Wu, J. (2006), \textit{{Scaling and Uncertainty Analysis in Ecology}}, Methods
  and Applications, Springer Science {\&} Business Media, Dordrecht, The
  Netherlands.

\bibitem[{\textit{Zhou et~al.}(2014)\textit{Zhou, Tian, Myneni, Ciais, Saatchi,
  Liu, Piao, Chen, Vermote, Song, and Hwang}}]{Zhou:2014gl}
Zhou, L., Y.~Tian, R.~B. Myneni, P.~Ciais, S.~Saatchi, Y.~Y. Liu, S.~Piao,
  H.~Chen, E.~F. Vermote, C.~Song, and T.~Hwang (2014), {Widespread decline of
  Congo rainforest greenness in the past decade}, \textit{Nature},
  \textit{509}(7498), 86--90.

\end{thebibliography}
%%%
%\bibliographystyle{agufull08}

------------------------------------------------------------------------ %%
%
%  END ARTICLE
%
%% ------------------------------------------------------------------------ %%
\end{article}
%
%%%%%%%%% Figures  %%%%%%%%%%%%%%%%%%%%%%%%%
\newpage
%%%% Figure 1
\newpage

%%%% Figure 2
\newpage 

%%%% Figures 3-4 merged
\newpage

%%%% Figure 4
\newpage

%%%% Figure 5
\newpage 

%%%% Figure 6
\newpage

%
% DO NOT USE \psfrag or \subfigure commands.
%
% Figure captions go below the figure.
% Table titles go above tables; all other caption information
%  should be placed in footnotes below the table.
%
%----------------
% EXAMPLE FIGURE
%
 %\begin{figure}
 %\noindent\includegraphics[width=20pc]{samplefigure.eps}
 %\caption{Caption text here}
 %\label{figure_label}
 %\end{figure}
%
% ---------------
% EXAMPLE TABLE
%
%\begin{table}
%\caption{Time of the Transition Between Phase 1 and Phase 2\tablenotemark{a}}
%\centering
%\begin{tabular}{l c}
%\hline
% Run  & Time (min)  \\
%\hline
%  $l1$  & 260   \\
%  $l2$  & 300   \\
%  $l3$  & 340   \\
%  $h1$  & 270   \\
%  $h2$  & 250   \\
%  $h3$  & 380   \\
%  $r1$  & 370   \\
%  $r2$  & 390   \\
%\hline
%\end{tabular}
%\tablenotetext{a}{Footnote text here.}
%\end{table}

% See below for how to make sideways figures or tables.

\end{document}